\documentclass[prb,twocolumn,superscriptaddress,showpacs,floatfix]{revtex4}

\usepackage{graphicx}
\usepackage{amssymb}
\usepackage{amsfonts}
\usepackage{bbold}
\usepackage{amsmath}
\usepackage{bm}

\usepackage{datetime}
\usepackage{color}

\begin{document}
\newcommand{\figdir}{.}
\newcommand{\figwidth}{0.9\columnwidth}
\newcommand{\ffigwidth}{0.4\columnwidth}
\newcommand{\bep} {\boldsymbol{\epsilon}}
\newcommand{\bc} {\mathbf{c}}
%
\title{Controlled engineering of extended states in disordered systems}

\author{Alberto Rodriguez}
\email[Corresponding author: ]{Alberto.Rodriguez.Gonzalez@physik.uni-freiburg.de}
\affiliation{
Physikalisches Institut, Albert-Ludwigs Universit\"{a}t Freiburg, Hermann-Herder Strasse 3, D-79104, Freiburg, Germany}
\author{Arunava Chakrabarti}
\affiliation{Department of Physics, University of Kalyani, Kalyani, West Bengal-741 235, India}
\author{Rudolf A. R\"omer}
\affiliation{
Department of Physics and Centre for Scientific Computing, University of Warwick, Coventry, CV4 7AL, United Kingdom}
\date{$Revision: 1.121 $, compiled \today, \currenttime}
%
\begin{abstract}
We describe how to engineer wavefunction delocalization in disordered systems modelled by tight-binding Hamiltonians in $d>1$ dimensions.
We show analytically that a simple product structure for the random onsite potential energies,
together with suitably chosen
hopping strengths, allows a resonant scattering process leading to ballistic transport along one direction, and 
a controlled coexistence of extended Bloch states and anisotropically localized states in the spectrum. 
We demonstrate that these features persist in the thermodynamic limit for a 
continuous range of the system parameters. Numerical results support these findings and highlight the robustness of the extended
regime with respect to deviations from the exact resonance condition for finite systems. 
The localization and transport properties of the system can be engineered almost at will and 
independently in each direction.
This study gives rise to the possibility of designing disordered potentials
that work as switching devices and band-pass filters for quantum waves, such as 
matter waves in optical lattices.
\end{abstract}

%
\maketitle

\section{Introduction}
\label{sec-introduction}

The electronic properties of most materials are determined by their
crystal structure or lack thereof. For purely crystalline materials,
well established methods of condensed matter physics \cite{AshM76}
allow the nearly complete experimental characterization, as well as
theoretical description, of the resulting electronic bands 
and associated density of states (DOS), all the way to transport and
thermal properties. 
In fact, our present understanding of the underlying mechanisms allows the manufacture of tailored
artificial crystal structures such as 
photonic,\cite{Yab87,Joh87} phononic,\cite{DeeJT98,VasDFH98} polaritonic\cite{BarAA09,GroP08} or plasmonic\cite{TaoSY07,ChrESG07} lattices.
While in the former two systems classical counterparts of electronic bands and
transport properties are manifestly observed, in the latter two,
electronic quasi-particle scattering is shown to lead to the formation of controllably engineered excitation
bands and, in particular, the \emph{gaps} between these as is required for
a multitude of applications.\cite{HepG11,Min11,RebWZI12}

In a strongly disordered system,
Anderson localization\cite{And58} suppresses transport even in regions
with a finite DOS.\cite{KraM93,EveM08} Particularly in low-dimensional systems,
the so-called scaling hypothesis\cite{AbrALR79} establishes the expectation
that all states remain localized for non-interacting quasi-particles, and
hence there seems little room for a similarly controlled "engineering" of
\emph{bands} of extended states. Nevertheless, such a complementary approach has
already enjoyed some successes.  
Local positional correlation in a disordered material 
has been shown to lead to resonant scattering
events generating extended states at isolated energies in the
spectrum.\cite{DunWP90,BelBHT99,ZhaU04,SedKS11} 
With much longer-ranged correlated disorder, 
even in one-dimensional (1D) systems, effective metal-insulator transitions can be induced.
\cite{AubA80,DemL98,IzrK99,KuhIKS00,MouCLR04,GuoX11}
Similarly, certain disordered configurations can lead to an optimization of the quantum interference process mediating excitonic transport in molecular networks. \cite{SchWB11,SchWB11b} 
In fact, a controlled disorder can induce highly selective transport properties,\cite{Hil03,XioKE98,MouD08,NdaRS04} giving rise to materials with interesting new functionalities, as has been recently explored with photonic crystals. \cite{AkaASN03,TonVOJ08,EstCBU12,GarSTL11}

\begin{figure*}
  \includegraphics[width=\columnwidth]{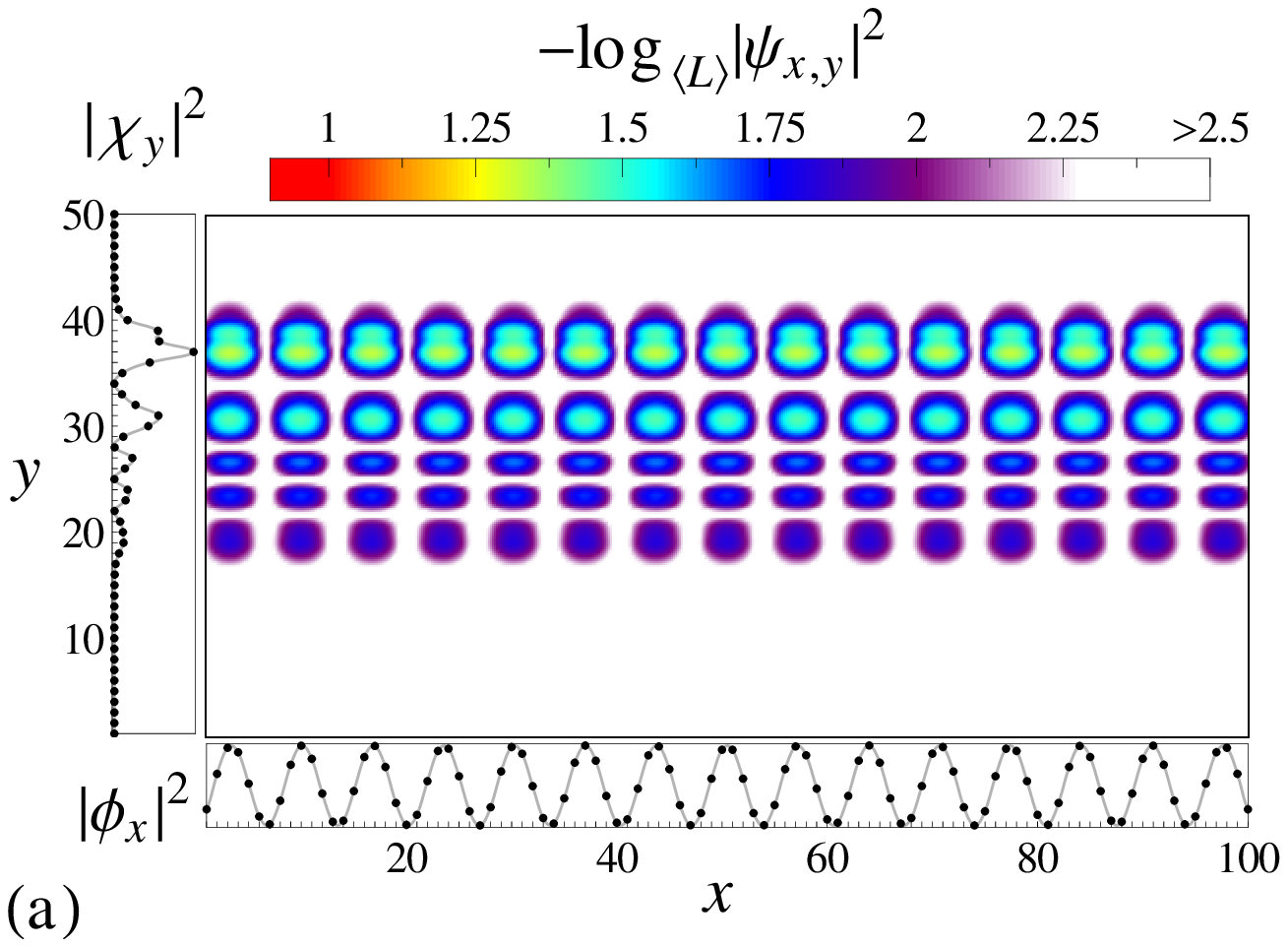}\hfill
  \includegraphics[width=\columnwidth]{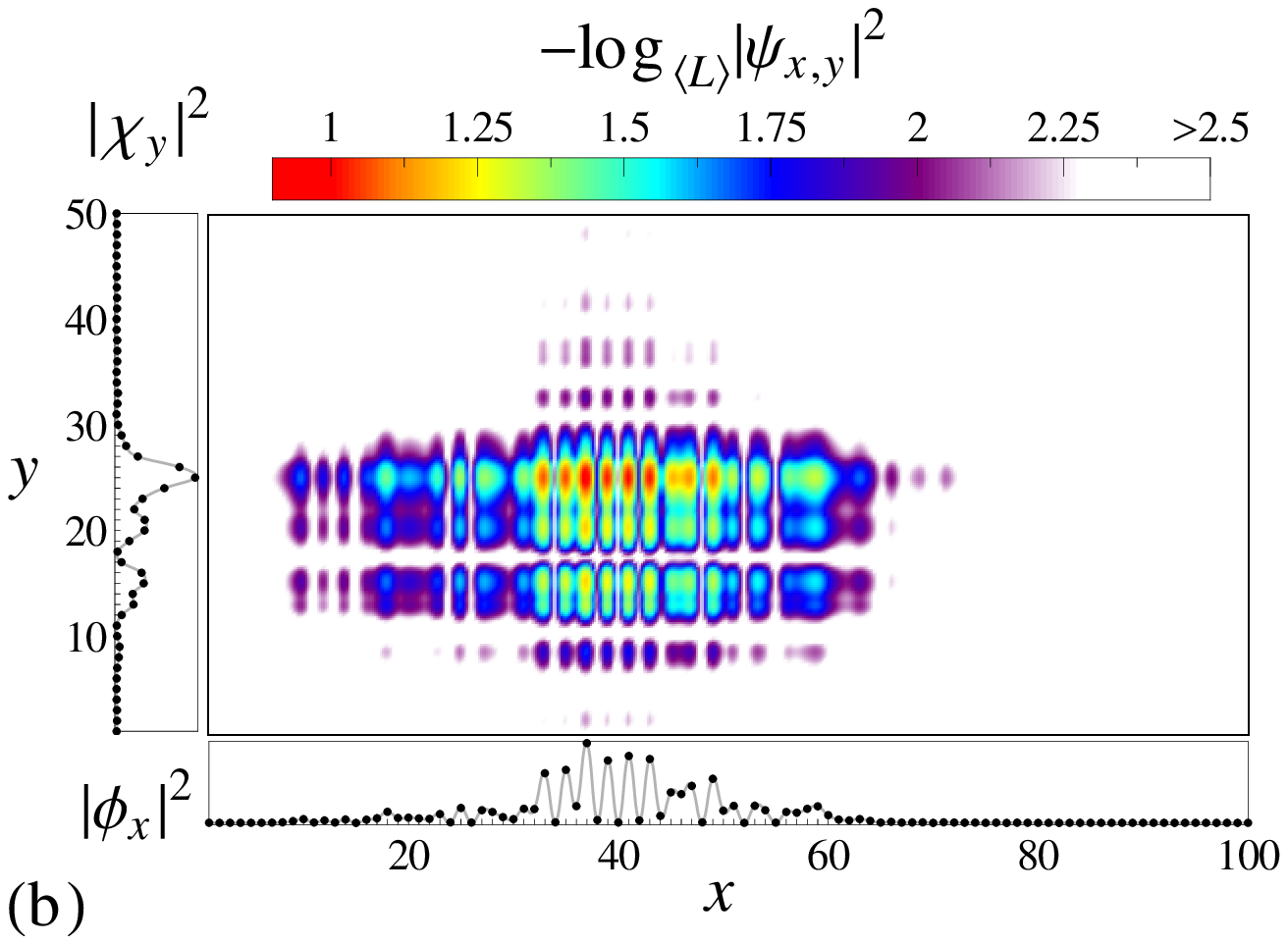}
 \caption{(Color online) Probability density $|\psi_{x,y}|^2$ of eigenstates of an $xy$-disordered system with (a) extended in $x$ and (b) localized behaviour. The parameters characterizing the potential in case (a) are $m_\alpha=0$, $W_\alpha=t$, $m_\beta=1$, $W_\beta=1$, $\xi_y\equiv\xi= 0.55488$ and eigenenergy $E/t= 1.7862$. For (b) we have $m_\alpha=0$, $W_\alpha=t$, $m_\beta=4$, $W_\beta=1$, $m_\xi=1$, $W_\xi=0.5$ and eigenenergy $E/t=-0.10995$. In both cases the system length and width are $L_x=100$ and $L_y=50$ respectively. The left and bottom panels show respectively the distributions $|\chi_y|^2$ and $|\phi_x|^2$ that construct the eigenstate according to \eqref{eq:2Dstate}. The color is determined by the value of $-\log_{\langle L\rangle}|\psi_{x,y}|^2$ where $\langle L\rangle\equiv\sqrt{L_xL_y}$ (in this scale, 2 is the average value of an extended state). A second order interpolation of the distributions is used to smooth the visualization.}
 \label{fig-states}
\end{figure*}

In this paper we show how to open a channel of perfect transmission in an otherwise disordered system.
Our approach uses a simple product structure for the random potential which, except for certain
resonance conditions, leads to anisotropically localized states. 
We believe our results to be of particular relevance for
ultracold atoms or Bose-Einstein condensates 
in optical lattices as these are ideal for studying disorder effects.\cite{BilJZB08,RoaDFF08,SanL10}
In fact, very recently the first direct observations of localization of matter waves in 3D disordered 
optical potentials have been reported. \cite{KonMZD11,JenBMC12}
Here, by engineering the underlying disorder, our study shows how
extended matter waves emerge from a background of localized states (cp.\ Fig.\ \ref{fig-states}).
Furthermore, our disorder structure allows for independent tailoring of the transport properties of the system
in each direction, and gives rise to energy coexistence of extended and localized states, which can be 
manipulated with a high degree of control. 
We prove our findings analytically and we corroborate them by extensive numerical studies. 

In Section \ref{sec-locspec}, we describe the chosen $xy$-disorder, restricting ourselves to 2D for presentational simplicity. 
We then deduce the form of the resulting DOS and show that the eigenstates exhibit localization. 
In Section \ref{sec-deloca}, we discuss the conditions for the existence of extended states and show the appearance of a perfectly transmitting channel.
Section \ref{sec-engineering} presents a discussion on the engineering of the transport properties and spectral regions with mixed localized and extended states.
In Section \ref{sec-conclusions} we draw our conclusions. 
Details on numerical techniques and some lengthy derivations are provided in several Appendices.

\section{Localization and spectral properties in $xy$-disorder}
\label{sec-locspec}

\subsection{The model}

\begin{figure*}
\includegraphics[width=.7\columnwidth]{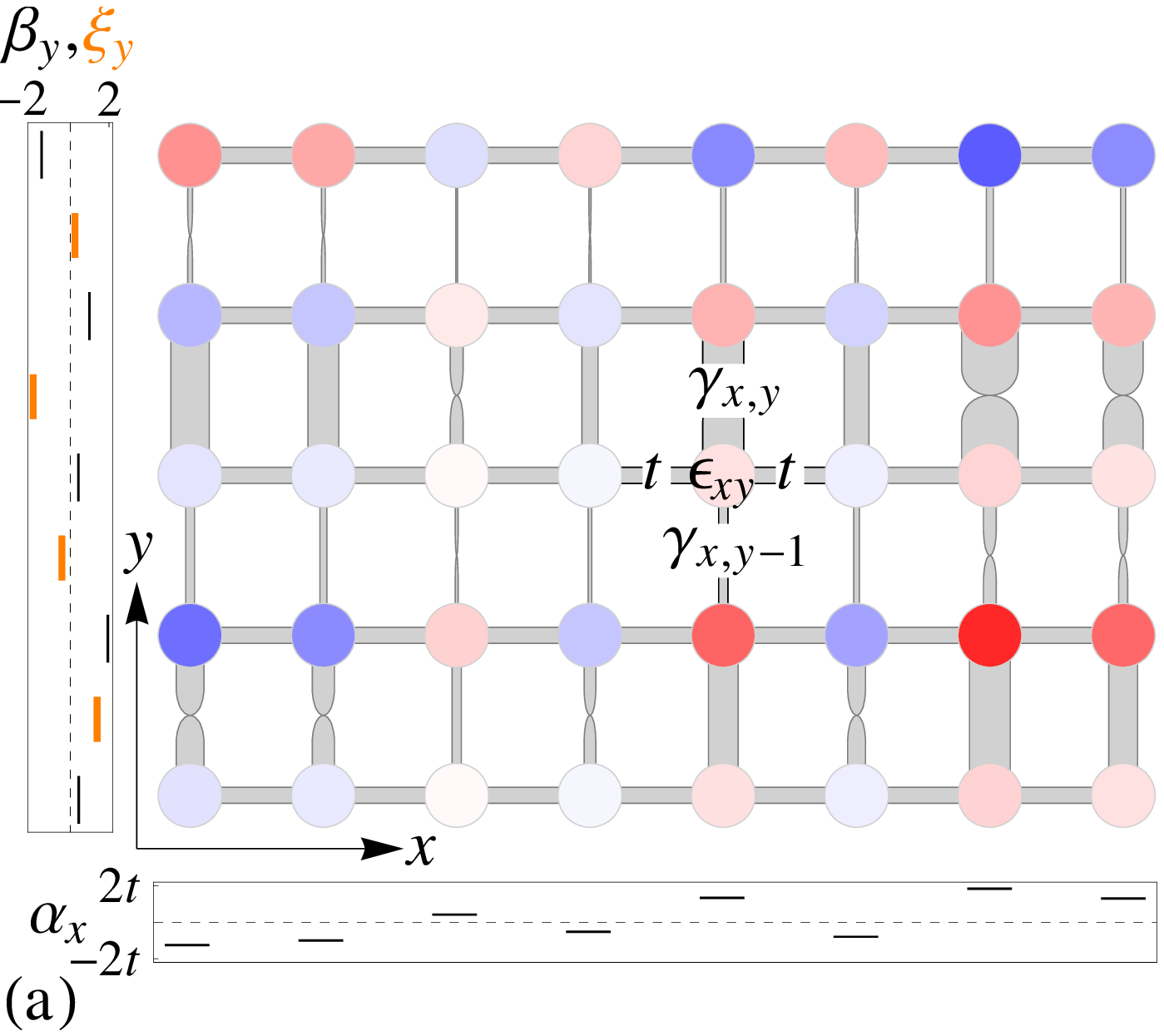}\hfill
\includegraphics[width=.65\columnwidth]{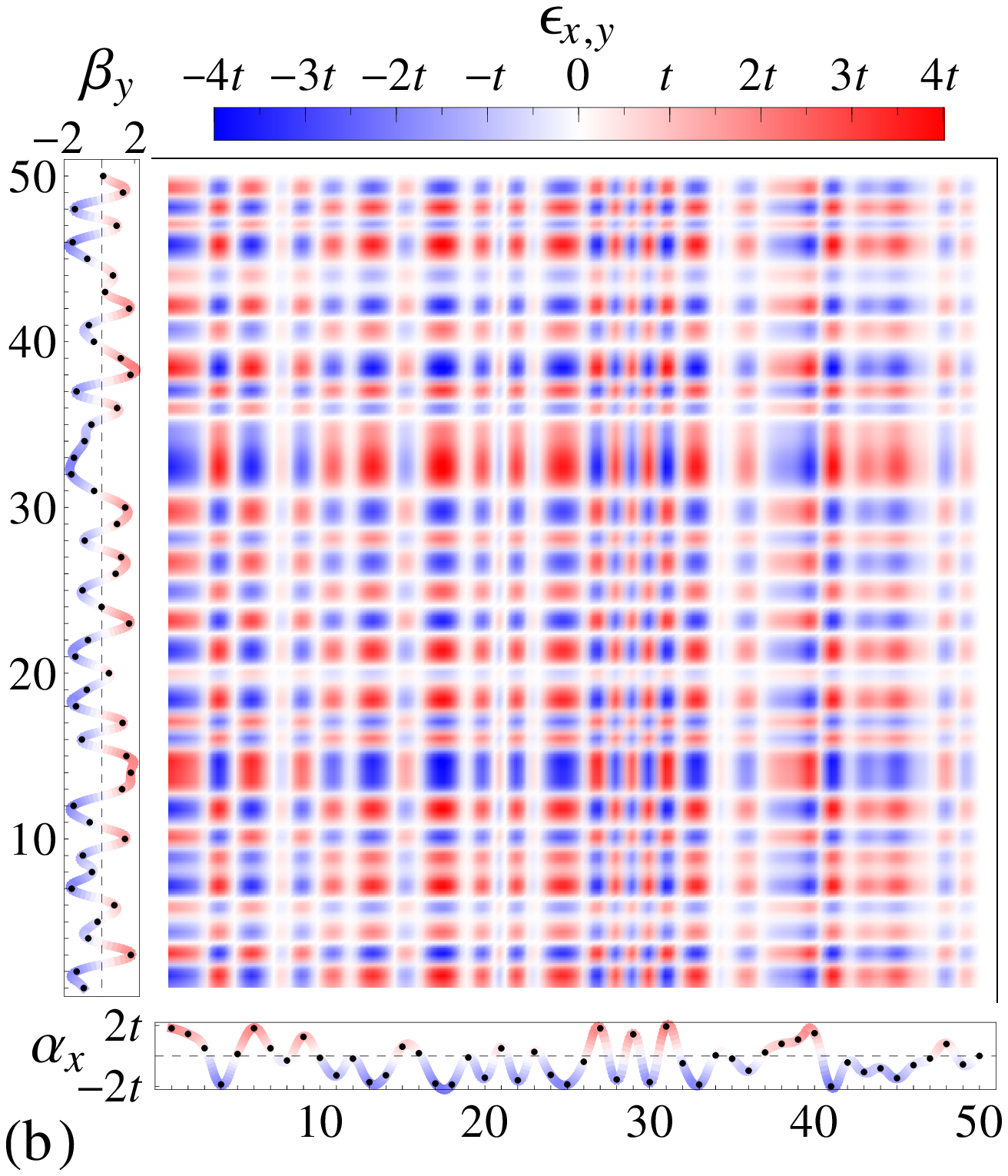}\hfill
\includegraphics[width=.65\columnwidth]{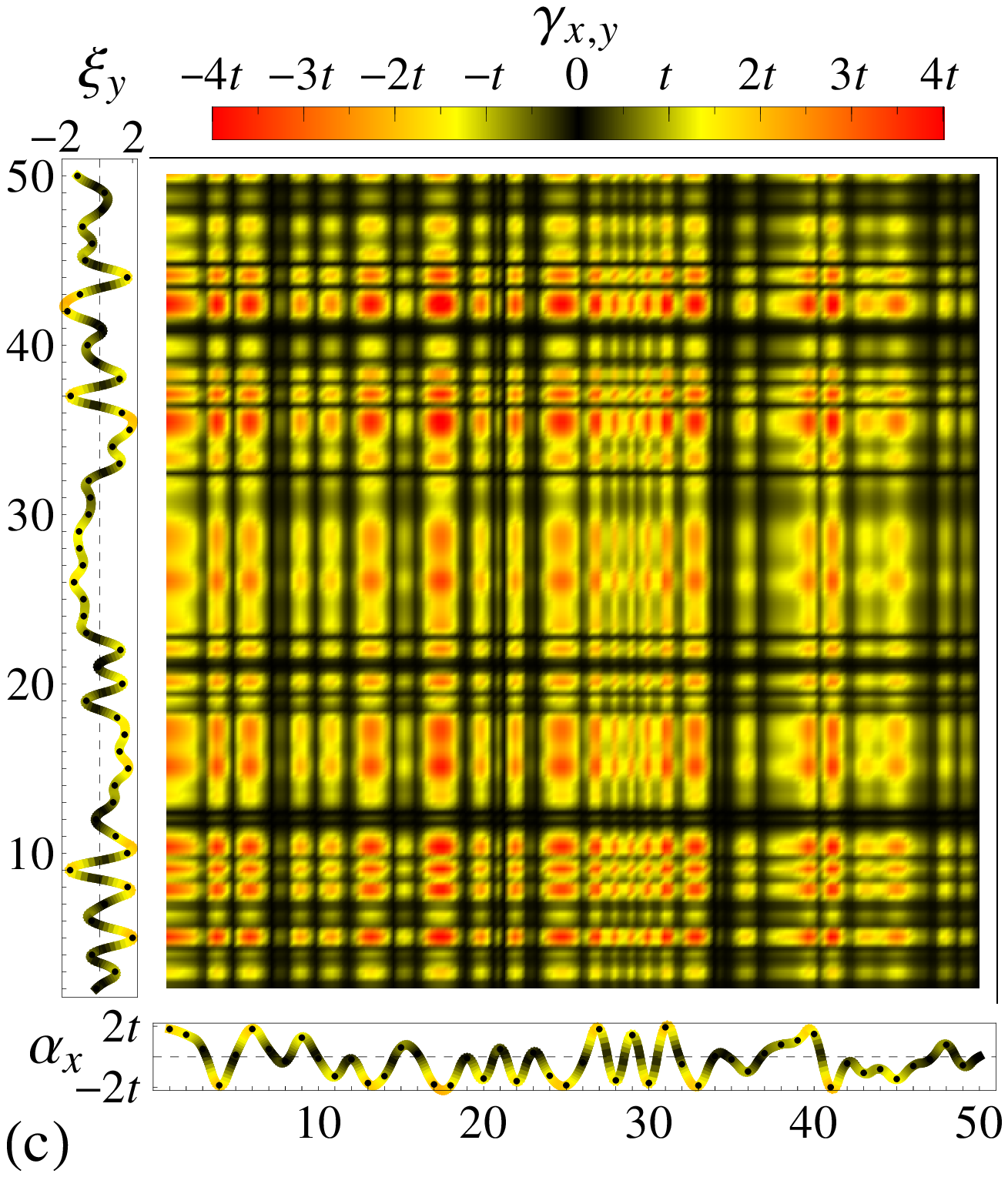}
 \caption{(Color online) 2D lattice with $xy$-disorder: (a) microscopic link-node representation. Different colors highlight different on-site energies at every site. The width of the links is proportional to the magnitude of the hopping terms. A twist in the link indicates a negative hopping term. The horizontal ($x$) and vertical ($y$) diagrams show the sequences $\{\alpha_x\}$ and $\{\beta_y\}$, $\{\xi_y\}$ respectively. Density plots for the on-site energies $\epsilon_{x,y}=\alpha_x\beta_y$ and the vertical hopping terms $\gamma_{x,y}=\alpha_x\xi_y$ for a $50\times50$ lattice with $m_\alpha/t=m_\beta=m_\xi=0$ and $W_\alpha/t=W_\beta=W_\xi=4$ are shown in (b) and (c) respectively. A second-order interpolation of the random sequences is used in (b) and (c) to smooth the visualization.}
 \label{fig-lattice}
\end{figure*}

We work with a 2D tight-binding model described by the Hamiltonian
\begin{equation}
 \mathcal{H}=\sum_{x=1}^{L_x}  \bc^\dagger_x \bep_x \bc_x + \sum_{x=1}^{L_x-1} (\bc^\dagger_x \mathbf{t} \bc_{x+1} +\bc^\dagger_{x+1} \mathbf{t}\bc_x),
\label{eq-model}
\end{equation}
on a lattice with $L_x \times L_y$ sites. Here, $\bep_x$ denotes the $L_y\times L_y$ Hamiltonian matrix acting in the $y$ direction for each vertical arm at position $x$ such that
\begin{equation}
 \bep_x\equiv\begin{pmatrix}
         \epsilon_{x,1} &\gamma_{x,1}  & 0 & \cdots & \cdots & 0 \\
	 \gamma_{x,1} & \epsilon_{x,2} & \gamma_{x,2} & 0 &   & \vdots \\
	 0 & \gamma_{x,2} & \epsilon_{x,3} & \gamma_{x,3} & 0 & \vdots  \\
         \vdots & 0 & \ddots & \ddots & \ddots & 0 \\
	\vdots  &   & 0 & \gamma_{x,L_y-2} & \epsilon_{x,L_y-1} & \gamma_{x,L_y-1} \\
 	0 & \cdots & \cdots & 0 & \gamma_{x,L_y-1} & \epsilon_{x,L_y}
        \end{pmatrix},
        \label{eq-bep}
 \end{equation}
 and $\bc_x^\dagger \equiv \left( c_{x,1}^\dagger, c_{x,2}^\dagger,  \ldots,  c_{x,L_y}^\dagger \right)$, with
$c_{x,y}$ ($c^\dagger_{x,y}$) the usual annihilation (creation) operators at the site with coordinates $\{x,y\}$. Also,  $\mathbf{t}\equiv t\,\mathbb{1}$ is the hopping along the $x$ direction.
The set $\{\epsilon_{x,y}\}$ gives the on-site energies and $\gamma_{x,y}$ is the hopping term in the $y$-direction between sites $(x,y)$ and $(x,y+1)$.
A pictorial representation of the lattice is shown in Fig.~\ref{fig-lattice}(a).

The energies and states of $\mathcal{H}$ are the solutions of the Schr\"{o}dinger equation
\begin{equation}
(E\mathbb{1}-\bep_x) \boldsymbol{\Psi}_x = t (\boldsymbol{\Psi}_{x+1} + \boldsymbol{\Psi}_{x-1}),
\label{eq:canonical}
\end{equation}
where $\boldsymbol{\Psi}_x \equiv \left( \psi_{x,1}, \psi_{x,2}, \ldots, \psi_{x,L_y}\right)^{\rm T}$ contains the wavefunction amplitudes in the $x$-th vertical arm of the system and $E$ is the energy. Here and in the following, we shall assume hard wall (fixed) boundary conditions in $x$ and $y$ direction, although the formalism does of course work with periodic boundaries as well.

\subsection{The disorder}

We now introduce disorder into the system in the following way. Let $\{\alpha_x\}_{x=1,\ldots,L_x}$ and $\{\beta_y\}_{y=1,\ldots,L_y}$
be random, uncorrelated sequences with corresponding probability distributions $\mathcal{P}(\alpha)$ and $\mathcal{P}(\beta)$ which for simplicity will be taken as box-distributions of widths $W_\alpha$, $W_\beta$, and mean values $\langle \alpha \rangle=m_{\alpha}$,  $\langle \beta \rangle = m_{\beta}$. The on-site random energies will be constructed from the \emph{product} of these two sequences, i.e.\
\begin{equation}
\epsilon_{x,y}\equiv \alpha_x\beta_y.
\label{eq-disorder}
\end{equation}
Additionally, we also choose the hopping strengths in $y$ direction to be random and determined by
\begin{equation}
\gamma_{x,y}\equiv \alpha_x \xi_y,
\label{eq-gamma}
\end{equation}
where $\{\xi_y\}_{y=1,\ldots,L_y-1}$ is another {\em independent} random sequence.
For simplicity, we choose the elements $\{\alpha_x\}$ to have dimensions of energy --- measured in units of the hopping $t$ ---,
thus $\{\beta_y\}$ and $\{\xi_y\}$ will be dimensionless.
This choice of $\epsilon_{x,y}$ and $\gamma_{x,y}$ above gives rise to patterns for the disordered energy landscape and the random (vertical) $y$ couplings with a characteristic correlation in the $x$ and $y$ directions, as can be seen in Figs.~\ref{fig-lattice}(b) and \ref{fig-lattice}(c), hence the name '$xy$-disorder'.
Let us remark that these choices, particularly for $L_y \ll L_x$, resemble and generalise quasi-1D ``ladder" models currently discussed in the literature to describe electronic transport in DNA \cite{KloRT05,ZhaYZD10} and mesoscopic devices.\cite{MouCL10,SilMC08,SilMC08b}
However we believe that a faithful realization of our model can be implemented using optical potentials and matter waves, where single-site resolution and control has already been demonstrated.\cite{OspOWS06,GreF08,BakGPF09,WurLGK09,SimBMT11,ShrWEC10,WeiESC11,BloDN12}

\subsection{Reduction to decoupled channels}

Equations \eqref{eq-disorder} and \eqref{eq-gamma} allow us to factorize the $\bep_x$ matrix as
\begin{eqnarray}
\bep_x &  = & \alpha_x\begin{pmatrix}
         \beta_1 &\xi_{1}  & 0 &\cdots & \cdots & 0 \\
	 \xi_{1} & \beta_2 & \xi_{2} & 0 &  & \vdots \\
	 0 & \xi_{2} & \beta_3 & \xi_{3} & 0 & \vdots  \\
         \vdots & 0 & \ddots & \ddots & \ddots & 0 \\
	 \vdots &  & 0 & \xi_{L_y-2} & \beta_{L_y-1} & \xi_{L_y-1} \\
 	0 & \cdots & \cdots & 0 & \xi_{L_y-1} & \beta_{L_y}
        \end{pmatrix} \notag \\
        & \equiv & \alpha_x \mathbf{P},
\label{eq:p}
\end{eqnarray}
where $\mathbf{P}$ does not depend on the $x$ coordinate.
The matrix $\mathbf{P}$ can be diagonalised via $\boldsymbol{p}=\mathbf{U}^{-1}\mathbf{PU}$, where $\mathbf{U}$ is the matrix whose columns are the orthonormal eigenvectors of $\mathbf{P}$ and $\boldsymbol{p}$ contains the eigenvalues $p_1, \ldots,p_{L_y}$ in its diagonal. Performing the change to a new basis
\begin{equation}
\boldsymbol{\Phi}_x=\mathbf{U}^{-1}\boldsymbol{\Psi}_x,
\label{eq:basischange}
\end{equation}
where $\boldsymbol{\Phi}_x \equiv \left( \phi_{x}^{(1)}, \phi_x^{(2)}, \ldots, \phi_x^{(L_y)}\right)^{\rm T}$, we can reduce Eq.~\eqref{eq:canonical} to
\begin{equation}
 (E\mathbb{1}-\alpha_x\boldsymbol{p}) \boldsymbol{\Phi}_x = t (\boldsymbol{\Phi}_{x+1} + \boldsymbol{\Phi}_{x-1}).
\label{eq:decoupled}
\end{equation}
This is simply a set of $L_y$ decoupled Schr\"{o}dinger equations,
\begin{align}
 (E-\alpha_x p_1) \phi_x^{(1)} =& \,t \left(\phi_{x+1}^{(1)} + \phi_{x-1}^{(1)}\right), \notag\\
		\vdots& \notag\\
 (E-\alpha_x p_c) \phi_x^{(c)} =& \,t \left(\phi_{x+1}^{(c)} + \phi_{x-1}^{(c)}\right), \label{eq-uncoupled}\\
	\vdots& \notag \\
 (E-\alpha_x p_{L_y}) \phi_x^{(L_y)} =& \,t \left(\phi_{x+1}^{(L_y)} + \phi_{x-1}^{(L_y)}\right), \notag
\end{align}
each of which corresponds to a 1D disordered channel with random on-site energies $\varepsilon_x^{(c)}\equiv \alpha_x p_c$ for the $c$-th {\em channel}. The disorder is uncorrelated within each channel and the distribution is $\mathcal{P}(\varepsilon)=\mathcal{P}(\alpha)/p_c$, which is again a box-distribution with mean $p_c m_\alpha$ and width $|p_c| W_\alpha$, for the $c$-th case.

\subsection{Density of states}
\label{sec-dos}
From Eqs.~\eqref{eq-uncoupled}, it follows that the energy spectrum of the 2D $L_x\times L_y$ system can
be obtained from the union of the different spectra for the corresponding decoupled 1D channels.
The spectrum will be determined by the properties of the random distributions $\{\alpha_x\}$, $\{\beta_y\}$, and $\{\xi_y\}$.
For finite $L_y$ the DOS per site of the 2D system can be written as
\begin{multline}
  g^{\textrm{2D}}(E;m_{\alpha},W_{\alpha},m_{\beta},W_{\beta},\{\xi_y\})=\\ \frac{1}{L_y}\sum_{p_c} g (E; p_c m_{\alpha},|p_c| W_{\alpha}),
 \label{eq:dos}
\end{multline}
where $g(E; p_c m_{\alpha},|p_c| W_{\alpha})$ is the DOS per site of the $c$-th decoupled channel, characterized by a disorder distribution of mean $p_c m_{\alpha}$ and width $|p_c| W_{\alpha}$.
In the limit $L_y\rightarrow\infty$, Eq.~\eqref{eq:dos} becomes
\begin{multline}
 g^\textrm{2D}(E; m_{\alpha},W_{\alpha}, m_{\beta},W_{\beta},\{\xi_y\})=\\ \int \varrho(p; m_{\beta}, W_{\beta},\{\xi_y\})\,g(E; p\, m_{\alpha}, |p| W_{\alpha})\,dp,
 \label{eq:dosinf}
\end{multline}
where $\varrho(p; m_{\beta}, W_{\beta},\{\xi_y\})$ corresponds to the DOS per site for the eigenvalues of $\mathbf{P}$. Due to the nature of $xy$-disorder, the DOS of the 2D system follows from a convolution of 1D  distributions.

\begin{figure}
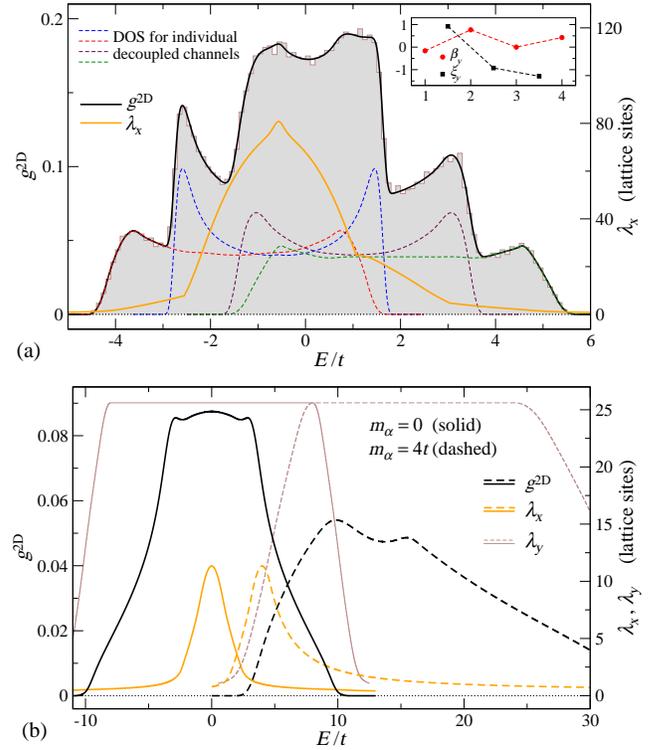

 \includegraphics[width=.96\columnwidth]{\figdir/fig-DOSLy4gen}\\[2mm]
 \includegraphics[width=.97\columnwidth]{\figdir/fig-DOSLyINFgen}
 \caption{(Color online) DOS $g^\text{2D}(E/t)$ and maximum localization lengths $\lambda(E)$ in the $x$ and $y$ directions for $xy$-systems as computed from the FEF. 
For panel (a) $L_x=\infty$ and $L_y=4$ for $m_\alpha=t$, $W_\alpha=2t$. The dashed lines show $g(E/t)/L_y$ for the $L_y$ decoupled infinite 1D channels that get added according to Eq.~\eqref{eq:dos} to produce $g^\text{2D}(E/t)$. For comparison, the shaded histogram has been obtained from exact diagonalization of a $(4\times1000)$-site system, averaging over 25 realizations of the $\alpha_x$ sequence. The inset shows the fixed finite sequences $\{\beta_y\}$ and $\{\xi_y\}$. 
 (b) $L_x=L_y=\infty$ with parameters $m_\beta=4$, $W_\beta=2$, $W_\alpha=3t$ and $m_\alpha=0$ (solid), $m_\alpha=4t$ (dashed). In this case $\xi_y\equiv\xi=1$. The DOS was calculated using Eq.~\eqref{eq:dosinf} where the integral over $p\in[1,7]$ was discretized using $1000$ points. Notice the different scale (right) for the localization lengths, obtained according to Eqs.\ \eqref{eq:locax} and \eqref{eq:locay}.}
 \label{fig-dosgeneric}
\end{figure}

Eqs.~\eqref{eq:dos} and \eqref{eq:dosinf} can be used to obtain the DOS numerically. Furthermore, since they only require calculations of 1-D distributions, they can be very efficiently combined with the functional equation formalism (FEF),\cite{Rod06} to obtain the DOS of the system for $L_x=\infty$ and either finite $L_y$ or $L_y=\infty$. A brief overview of the FEF is given in Appendix \ref{ap-FEF}. Some examples of the DOS for $xy$-disorder are shown in Fig.~\ref{fig-dosgeneric}, where we see 
that the global distribution of states is a superposition of the DOS from the individual channels.

\subsection{Localization of eigenstates}
\label{sec-localization}
The states of the 2D system can be obtained from the eigenstates of the 1D decoupled channels by undoing the change of basis \eqref{eq:basischange}, $\boldsymbol{\Psi}_x= \mathbf{U} \boldsymbol{\Phi}_x$. If we consider the $n$-th eigenenergy of the $c$-th decoupled channel with corresponding eigenstate $\{\phi^{(c,n)}_{x}\}_{x=1,\ldots,L_x}$, the vector  $\boldsymbol{\Phi}_x$ will only have one non-zero component at the $c$-th position. It then follows that
\begin{eqnarray}
 \boldsymbol{\Psi}_x
  &= &\mathbf{U} \cdot \begin{pmatrix} 0, \ldots, 0, \phi^{(c,n)}_{x}, 0, \ldots, 0\end{pmatrix}^{T} \notag\\
  &\equiv & \mathbf{U}^{(c)} \phi^{(c,n)}_x,
   \label{eq:BT}
\end{eqnarray}
where $\mathbf{U}^{(c)}$ means the $c$-th column of $\mathbf{U}$, which corresponds to the $c$-th eigenvector of the matrix $\mathbf{P}$ which we denote by $\{\chi^{(c)}_{y}\}_{y=1,\ldots,L_y}$. The 2D eigenstates of \eqref{eq-model} are thus obtained as
\begin{equation}
  \psi_{x,y}=\chi^{(c)}_{y}\phi^{(c,n)}_x, \qquad \begin{aligned} x &= 1,\ldots,L_x \\ y &=1,\ldots,L_y \end{aligned},
  \label{eq:2Dstate}
\end{equation}
where we must consider all 1D eigenstates $n=1,\ldots,L_x$ for all the decoupled channels $c=1,\ldots,L_y$, giving the complete basis of $L_x\times L_y$ states in the original coordinates. Notice that all eigenstates of a certain decoupled channel $c$ get multiplied by the same eigenvector of $\mathbf{P}$.

From Eqs.~\eqref{eq:p} and \eqref{eq:decoupled}, and making use of known
results on Anderson localization in 1D,\cite{AbrALR79} we conclude that, for {\em generic} cases, the eigenstates \eqref{eq:2Dstate} will be {\em anisotropically localized} for all energies, with different localization lengths in the $x$ and $y$ directions. Localization
persists although our model \eqref{eq-model} only contains $L_x+L_y-1$ independent random on-site energies  $\{\epsilon_{x,1}, \epsilon_{1,y}, \text{for } x=1, \ldots, L_x,y=2,\ldots,L_y\}$, as compared to the standard Anderson model with $L_x \times L_y$ uncorrelated values for $\epsilon_{x,y}$. An example of a localized eigenstate is shown in Fig.~\ref{fig-states}(b).

For a given energy $E$, the transport properties of the system in the $x$ direction, as $L_x\rightarrow\infty$, will be determined by the largest possible value of the localization length in this direction, $\lambda_x(E)$. It then follows that
\begin{equation}
  \lambda_x(E)=\max \left\{\lambda_x^{(c)}(E)\right\}_{c=1,\ldots,L_y},
  \label{eq:locax}
\end{equation}
where the localization length for each channel $c$ at energy $E$ is defined as the inverse of the corresponding Lyapunov exponent, $\lambda_x^{(c)}(E)\equiv1/\eta_x^{(c)}(E)$.\cite{KraM93}
Additionally, localization in the $y$ direction will be determined by the Lyapunov exponent of the 1D system defined by the $\mathbf{P}$ matrix only, $\lambda_y(p)\equiv 1/\eta_y(p)$, for large enough $L_y$. The localization length in $y$ will change with the eigenvalue $p$, and thus all wavefunctions \eqref{eq:2Dstate} obtained from the same decoupled channel [same $p_c$ in Eqs.~\eqref{eq:decoupled}] will exhibit the same $y$-spreading. The relevant localization length in $y$ as a function of the energy can be defined as
\begin{equation}
 \lambda_y(E)=\max \left\{\lambda_y(p_c)\right\}_{p_c / E\in S_c},
 \label{eq:locay}
\end{equation}
where
\begin{equation}
S_c\equiv\left[p_c\left(m_\alpha -\sigma \frac{W_\alpha}{2}\right)-2t, p_c\left(m_\alpha +\sigma \frac{W_\alpha}{2}\right)+2t\right]
\label{eq-channelspectrum}
\end{equation}
 denotes the boundaries of the energy spectrum of the $c$-th channel as $L_x\rightarrow\infty$, and $\sigma\equiv \text{sign}(p_c)$. That is, at energy $E$,  the relevant localization length in $y$ will be the maximum of $\lambda_y(p_c)$ over all channels whose spectrum includes $E$.

The Lyapunov exponents for the decoupled 1D channels, in the thermodynamic limit, can be obtained numerically using the FEF (see Appendix \ref{ap-FEF}), which in combination with \eqref{eq:locax} and \eqref{eq:locay} permits the calculation of the relevant localization lengths for a 2D system with generic $xy$-disorder, like those shown in Fig.~\ref{fig-dosgeneric}. We emphasize that, irrespective of the detailed energy dependence of these localization lengths, all $\lambda$ values are finite and hence correspond to localized states.

\begin{figure}
 \includegraphics[width=.95\columnwidth]{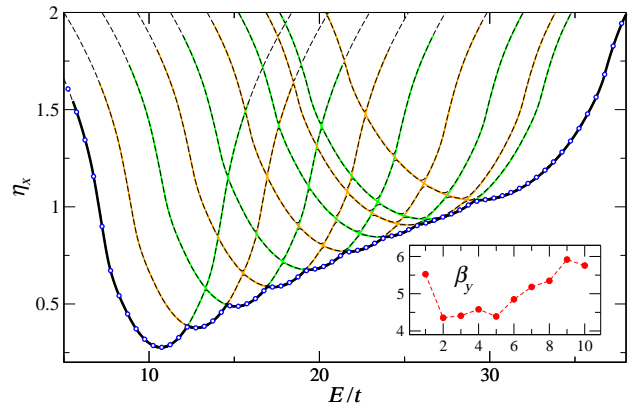}
 \caption{(Color online) Full spectrum of Lyapunov exponents $\eta_x(E)$ from TMM (dashed lines) and FEF (solid lines) for a system with $xy$-disorder ($L_x=\infty$, $L_y=10$) and parameters $m_\alpha=4t$, $W_\alpha=2t$, and $\xi_y\equiv\xi=1$. The inset shows the 10-element sequence considered for $\beta_y$. The thick black line highlights the trajectory of the minimum Lyapunov exponent for each energy, i.e. the inverse of the maximum localization length, as computed by the FEF. The blue circles show every 50th data point of the corresponding TMM result (errors are within symbol size).}
 \label{fig-TMMvsFEF}
\end{figure}

In order to confirm the validity of the calculation of the localization lengths, we carried out transport simulations in the $x$ direction  using the transfer-matrix method (TMM)\cite{KraM93} (see Appendix \ref{ap-tmm}) on Eq.\ \eqref{eq:canonical} without performing any decoupling. This technique is capable of giving the whole spectrum of characteristic Lyapunov exponents and thus automatically provides the largest decay (localization) length for an initial excitation travelling through the system.
In Fig.~\ref{fig-TMMvsFEF} we show the whole Lyapunov spectrum of an $xy$-disordered system ($L_x\rightarrow\infty$ and finite $L_y$) obtained from TMM, and compare it with the results from the FEF on the decoupled channels. The excellent agreement observed confirms the correctness of the decoupling transformation and of Eq.~\eqref{eq:locax}.

\section{Delocalization in $xy$-disorder}
\label{sec-deloca}
\begin{figure}[t]
 \includegraphics[width=.95\columnwidth]{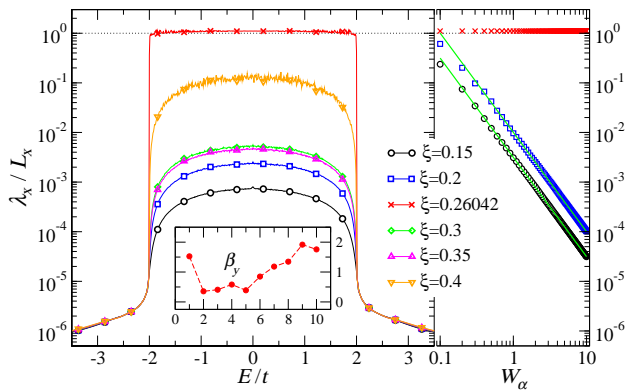}
   \caption{(Color online) Energy dependence of the normalized localization length in the $x$ direction obtained from TMM
   for a system of $L_y=10$, $m_{\alpha}=0$ and $W_{\alpha}=2t$
   for different values of $\xi_y\equiv\xi$.
   For clarity only every 50th data point has been indicated by a symbol. Errors are within symbol size.
   For the chosen realisation of $\beta_{y}$ values (shown in inset), $\xi=0.26042$ corresponds to $p_c=0$ for one of the $10$ channels.
   Note that the TMM calculations were terminated at a fixed value of $L_x=10^6$ and hence the
   $\lambda_x$ values for $\xi=0.26042$ in the energy range $|E|\leqslant 2$ do not diverge.
   The right frame shows the dependence of $\lambda_x/L_x$ versus $W_{\alpha}$ at energy $E=0$. The green lines represent $W_{\alpha}^{-2}$ power laws.
   }
  \label{fig-trans-E-allL}
\end{figure}

Despite the localized nature of eigenstates in generic $xy$-disorder, a {\em genuine delocalization} in the $x$-direction can be achieved as well. This is demonstrated in Fig.~\ref{fig-trans-E-allL},
where we show results from TMM calculations of the localization length $\lambda_x(E)$ for a system with $L_y=10$ and different values of $\xi_y\equiv\xi$ (here we assume a constant $\xi$ for presentational simplicity only).
We see that the enhancement of the localization length is always found inside the region $|E|\leqslant 2$.
In fact, as shown in Fig.~\ref{fig-trans-E-allL}, for certain $\xi$ the localization length becomes comparable to $L_x$ independently of the longitudinal disorder $W_{\alpha}$, signalling the appearance of extended states. This is in contrast to the $W_{\alpha}^{-2}$ dependence expected for quasi-1D systems,\cite{Eco90,RomS04} which is found at most other $\xi$ values.

\subsection{Zero eigenvalue condition}
\label{sec-zero}

The reason for the existence of this band of  extended states can be understood by returning to Eq.\ \eqref{eq-uncoupled}. We see that if at least one of the eigenvalues $\{p_c\}$ of $\mathbf{P}$ is zero, then the corresponding 1D equation in \eqref{eq-uncoupled} no longer describes localization but rather an extended channel. As each $p_c$ value depends on the distributions $\{\beta_y\}$ and $\{\xi_y\}$, the condition $p_c=0$ can be controlled by fine-tuning these parameters in our 2D disordered system, therefore inducing {\em a perfectly transmitting channel}.

Indeed, the relation between the $\{\xi_y\}$ and $\{\beta_y\}$ values can be derived straightforwardly. The characteristic polynomial of $\mathbf{P}$ is
\begin{equation}
 p^{L_y} - p^{L_y-1} \text{tr}\mathbf{P} +\cdots + f(\beta_y,\xi_y) p +(-1)^{L_y} \det\mathbf{P} = 0,
\end{equation}
and if $\det\mathbf{P}(\beta_y,\xi_y)=0$ then this will correspond to at least one eigenvalue of $\mathbf{P}$ being zero. Assuming that the $\beta_y$ disorders have been fixed, the condition can be satisfied by choosing the remaining $\xi_y$ values appropriately.
For example, for $L_y=2$, $3$, $4$, the condition $p_c=0$ reads
\begin{align}
 \xi_1^2 &= \beta_1\beta_2, \\
 \xi_1^2+\xi_2^2 &= \frac{\beta_1\beta_2\beta_3}{\beta_1+\beta_3}, \\
 \xi_1^2\xi_3^2 &= \beta_1\beta_2\xi_3^2+\beta_1\beta_4\xi_2^2+\beta_3\beta_4\xi_1^2-\beta_1\beta_2\beta_3\beta_4,
\end{align}
respectively.
In the most general case one could select arbitrarily $L_y-2$ values for $\xi_y$ and then choose the remaining one such that $\det\mathbf{P}=0$. From here on, we shall restrict ourselves to the case where $\xi_y\equiv \xi$ constant for all $y$. This simplification makes the hoppings in the $y$ direction constant within each vertical arm, i.e.\ $\gamma_{x,y}\equiv \alpha_x \xi$. It should be clear, however, that this condition is in general not necessary for the existence of extended states.

Because of the tridiagonal nature of $\mathbf{P}$, its determinant involves only even powers of $\xi$, and it is a polynomial of order $L_y$ [$(L_y-1)$] for  even [odd] $L_y$. We can then consider only $\xi>0$ without loss of generality.
Therefore, the spectrum of $\xi$ satisfying $\det\mathbf{P}(\xi)=0$ contains at most $L_y/2$ or $(L_y-1)/2$ independent positive real values.
\cite{foot-realsolution}
For any of these resonant $\xi$ values \cite{foot-precision} 
($p_c=0$), the spectrum of the otherwise disordered system has at least one perfectly conducting channel populated by $L_x$ extended states in the $x$ direction, in the energy interval $E\in[-2t,2t]$.

\subsection{Spectrum and stability of resonant channel}
\label{sec-stability}
\begin{figure}
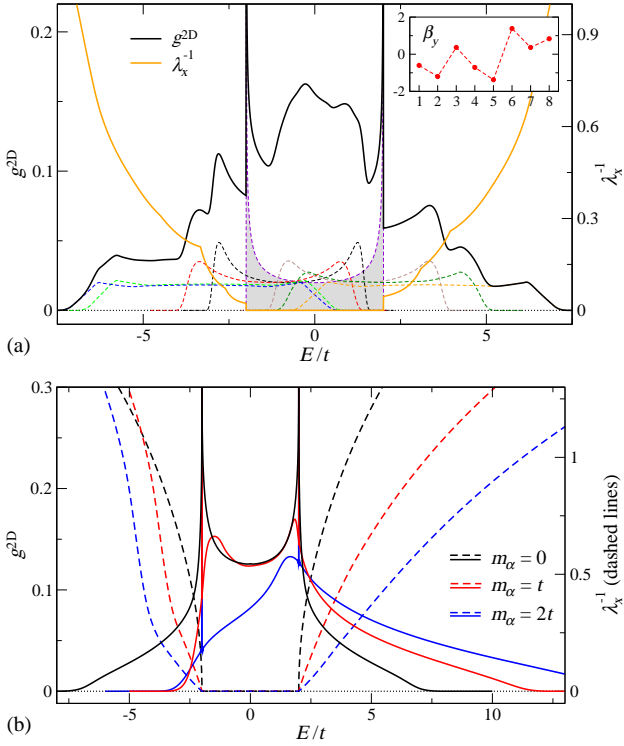

 \includegraphics[width=.95\columnwidth]{\figdir/fig-DOSLy8res}\\[2mm]
 \includegraphics[width=.95\columnwidth]{\figdir/fig-DOSLyINFres}
 \caption{(Color online) DOS and inverse localization length in the $x$ direction for $xy$-disorder with resonant channel, obtained from FEF.
 (a) $L_x=\infty$ and $L_y=8$ for $m_\alpha=2t$, $W_\alpha=3t$ and $\xi=0.44380$. The dashed lines show $g(E/t)/L_y$ for the decoupled infinite 1D channels, that get added according to Eq.~\eqref{eq:dos} to produce $g^\text{2D}(E/t)$. The resonant channel contribution is highlighted in gray. The inset shows the fixed finite sequence $\{\beta_y\}$. 
 (b) $L_y=L_x=\infty$ with parameters $m_\beta=2$, $W_\beta=2$, $W_\alpha=3t$ and different $m_\alpha$.  In this case $\xi=1$. DOS was calculated using Eq.~\eqref{eq:dosinf} where the integral over $p\in[-1,5]$ was discretized using 1000 points. Notice the different scale (right) for the inverse localization lengths.}
 \label{fig-dosresonant}
\end{figure}

The eigenstates of the resonant channel [with $p_c=0$ in Eq.~\eqref{eq:decoupled}] are Bloch states, \cite{foot-bloch} 
\begin{equation}
|\phi^{(c,n)}_x|^2=\frac{2}{L_x+1}\sin^2\left(\frac{n\pi}{L_x+1}x\right),
\label{eq-bloch}
\end{equation}
with energy $E^{(n)}=2 t \cos\left[\pi n/(L_x+1)\right]$, $n=1,\ldots, L_x$, for hard-wall boundary conditions $\phi^{(c,n)}_0=\phi^{(c,n)}_{L+1}=0$.
However, the 2D eigenstates given by Eq.~\eqref{eq:2Dstate}, are in general still localized in the $y$ direction. An example of such states is shown in Fig.~\ref{fig-states}(a).
The presence of these extended states changes also the properties of the DOS of the system. According to Eq.~\eqref{eq:dos}, the DOS now contains the contribution from the spectrum of a periodic chain, $g(E)=\pi^{-1}(4t^2-E^2)^{-1/2}$ for $E\in[-2t,2t]$,  and its characteristic Van-Hove singularities at the band edges, as shown in Fig.~\ref{fig-dosresonant}(a).

\begin{figure}
 \includegraphics[width=.95\columnwidth]{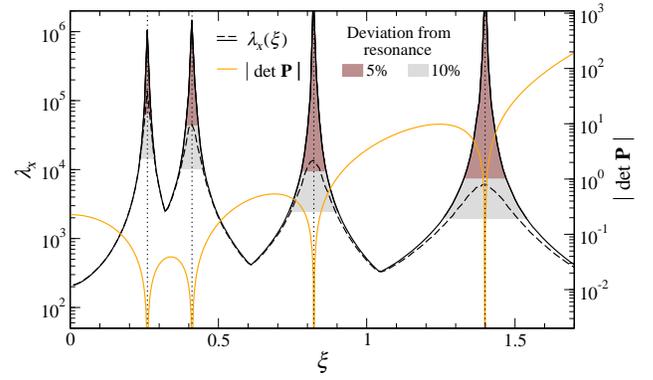}
 \caption{(Color online) Evolution of $\det\mathbf{P}$ (right scale) and localization length $\lambda_x (E=0)$ (left scale, black solid and dashed lines) as a function of $\xi$ for the $xy$-system used in Fig.~\ref{fig-trans-E-allL}. Four resonant values appear in the interval shown $\xi=0.26042$, $0.41092$, $0.82164$, and $1.2992$. The shaded regions highlight the intervals of $\xi$ corresponding to $\pm 5\%$ and $\pm10\%$ deviation around the resonances. The dashed line corresponds to calculations of $\lambda_x (E=0)$ where each $\gamma_{x,y}$ value was allowed to deviate randomly up to $\pm 5\%$ from the target $\gamma_{x,y}=\alpha_{x}\xi$ value. }
 \label{fig-stability}
\end{figure}

In general, there is a clear correlation between the dependences of $\det\mathbf{P}$ and $\lambda_x$ on $\xi$, as shown in Fig.~\ref{fig-stability}. Therefore very large localization lengths occur in the vicinity of the resonant $\xi$ values, which will lead to {effectively extended} channels in finite systems, even if the resonant condition is not precisely met. For the case considered in Fig.~\ref{fig-stability}, corresponding to a quasi-1D system with $L_y=10$, we see that deviations from the resonant $\xi$ value up to $\pm 10\%$ still ensure localization lengths in the $x$ direction of at least several thousands of lattice sites. More interestingly, we have also checked that $5\%$ of random spatial deviations in each $\gamma_{x,y}$ value --- i.e. when the factorization \eqref{eq-gamma} for the vertical hoppings is weakly broken ---, do have a similar effect only. This tolerance of the delocalization channel is very significant, and makes our results relevant for potential experimental realizations, where deviations from the theoretically obtained parameters are to be expected. Furthermore, the situation becomes even more robust when the width of the system is increased, as we discuss below.
%

\subsection{The resonant condition for large $L_y$}
\label{sec-largesize}

The resonant condition requires $p=0$ to be an eigenvalue of the matrix $\mathbf{P}$, defined in Eq.~\eqref{eq:p}. For short $L_y$ this implies tuning the vertical hopping elements via $\xi$, for which several isolated values lead to the appearance of a transmitting channel. For large $L_y$, however, the resonant condition translates into whether the DOS for the system described by $\mathbf{P}$ satisfies
$\varrho(p=0)\neq 0$. If the latter is true, then $p=0$ is an eigenvalue as $L_y\rightarrow\infty$, and
at least one of the decoupled channels of the system will not correspond to a disordered chain.
Therefore the infinite system ($L_y=\infty$, $L_x=\infty$) will always exhibit
a perfect transmission band in the $x$ direction for $E\in[-2t,2t]$ as long as $\varrho(p=0)\neq0$.

The matrix $\mathbf{P}$ has the structure of a 1D Anderson model with disordered onsite potentials, $\beta_y$, constant hopping strength, $\xi_y\equiv\xi$, and hardwall boundaries. By Gershgorin's circle theorem,\cite{Ger31,GolL96} the spectrum of eigenvalues $p$ as $L_y\rightarrow\infty$ fills the interval $p\in[-2 |\xi| + m_{\beta} - W_{\beta}/2,  m_{\beta} + W_{\beta}/2 +2|\xi|]$. The condition for the existence of a perfect conducting channel is then
\begin{equation}
 |\xi| > \frac{|m_{\beta}|}{2} - \frac{W_{\beta}}{4}.
 \label{eq:resoTL}
\end{equation}
The above inequality can be satisfied either by changing $\xi$ or the $\beta_y$ distribution. As an example, we see that when $m_\beta=0$, the resonant channel will emerge (for large $L_y$) for any value of $\xi$.
Numerical results for DOS and localization length for  $xy$-systems in the thermodynamic limit ($L_x=L_y=\infty$) when the above inequality is fulfilled are shown in Fig.~\ref{fig-dosresonant}(b).
We emphasize that for large $L_y$, due to the nature of the DOS in the 1D Anderson model (smooth and differentiable), the condition $\varrho(p=0)\neq0$ implies that the fraction of values of $p$ arbitrarily close to $0$ is finite, and therefore the number of channels with very large localization lengths grows with $L_y$. These will then lead to a dense continuum of effectively extended states for $E\in[-2t,2t]$ in systems with large but finite $L_x$ and $L_y$ whenever the above condition is satisfied.

We can estimate the relation between $\lambda_x$ and $L_y$ for nearly resonant channels, i.e. when \eqref{eq:resoTL} holds. If we assume that $\varrho(p)\equiv\varrho$ is roughly constant around $p=0$, then for finite $L_y$ we will find a value of $p$ as small as
$|p_s|\leqslant(2L_y\varrho)^{-1}$. The localization in the corresponding decoupled channel --- which provides the maximum localization length ---, characterized by a disorder distribution of width $|p_s| W_\alpha$ and mean $p_s m_\alpha$, at $E=0$ is given by (cp.\ Fig.\ \ref{fig-trans-E-allL})
\begin{equation}
 \lambda_x(E=0)= \frac{24}{p_s^2 W_\alpha^2} \left[4t^2-p_s^2 m_\alpha^2\right],
\end{equation}
since $|p_s| W_\alpha$ is small. \cite{Tho72,RomS04}
This leads to
\begin{equation}
 \lambda_x(E=0)\geqslant \frac{384 L_y^2 \varrho^2}{(W_\alpha/t)^2} -\frac{24 m_\alpha^2}{W_\alpha^2}.
\end{equation}
Therefore the minimum localization length at the band centre of the nearly resonant channels, which will emerge when Eq.~\eqref{eq:resoTL} is fulfilled, scales as $L_y^2$. We can roughly estimate the order of magnitude of $\varrho$ by assuming a constant DOS which leads to $\varrho=(W_\beta+4|\xi|)^{-1}$. For the typical parameters used in our simulations,
we then obtain $\lambda_x(E=0) > 3 L_y^2$ as approximate lower bound. We note that this estimate for large $L_y$ already works quite well for the $L_y=10$ case of Fig.\ \ref{fig-stability}.

\subsection{Effective delocalization in $y$}
\label{sec-ydeloc}
In general, for large enough $L_y$, localization of the wavefunctions given in \eqref{eq:2Dstate} is to be expected in the $y$-direction, with localization lengths determined by the eigenvectors of $\mathbf{P}$ (cp.\ Fig.\ \ref{fig-states}).
The properties of the 1D disordered system described by the latter matrix depend on $\xi$ and $\beta_y$, corresponding, respectively, to hopping and on-site energies. Hence we can write the localization length in the $y$-direction as
\begin{equation}
 \lambda_y(p)= \frac{24}{W_\beta^2} \left[4\xi^2-\left(p -m_\beta\right)^2\right],
 \label{eq:lly}
\end{equation}
in the weak disorder approximation, $|\xi|\gg W_\beta/\sqrt{12}$.\cite{Tho72,RomS04} As explained in Section \ref{sec-localization}, the transverse localization is determined by the value of $p$, and thus all states  of a certain decoupled channel characterized by $p_c$ [see Eqs.~\eqref{eq:decoupled}] will exhibit the same $y$-spreading. Localization in $y$ can then also be tailored by changing $\xi$ and/or the disorder distribution $\beta_y$.
For example, for fixed disorder, $\lambda_y$ can be increased by choosing larger $\xi$ values.

\begin{figure*}
  \includegraphics[width=\columnwidth]{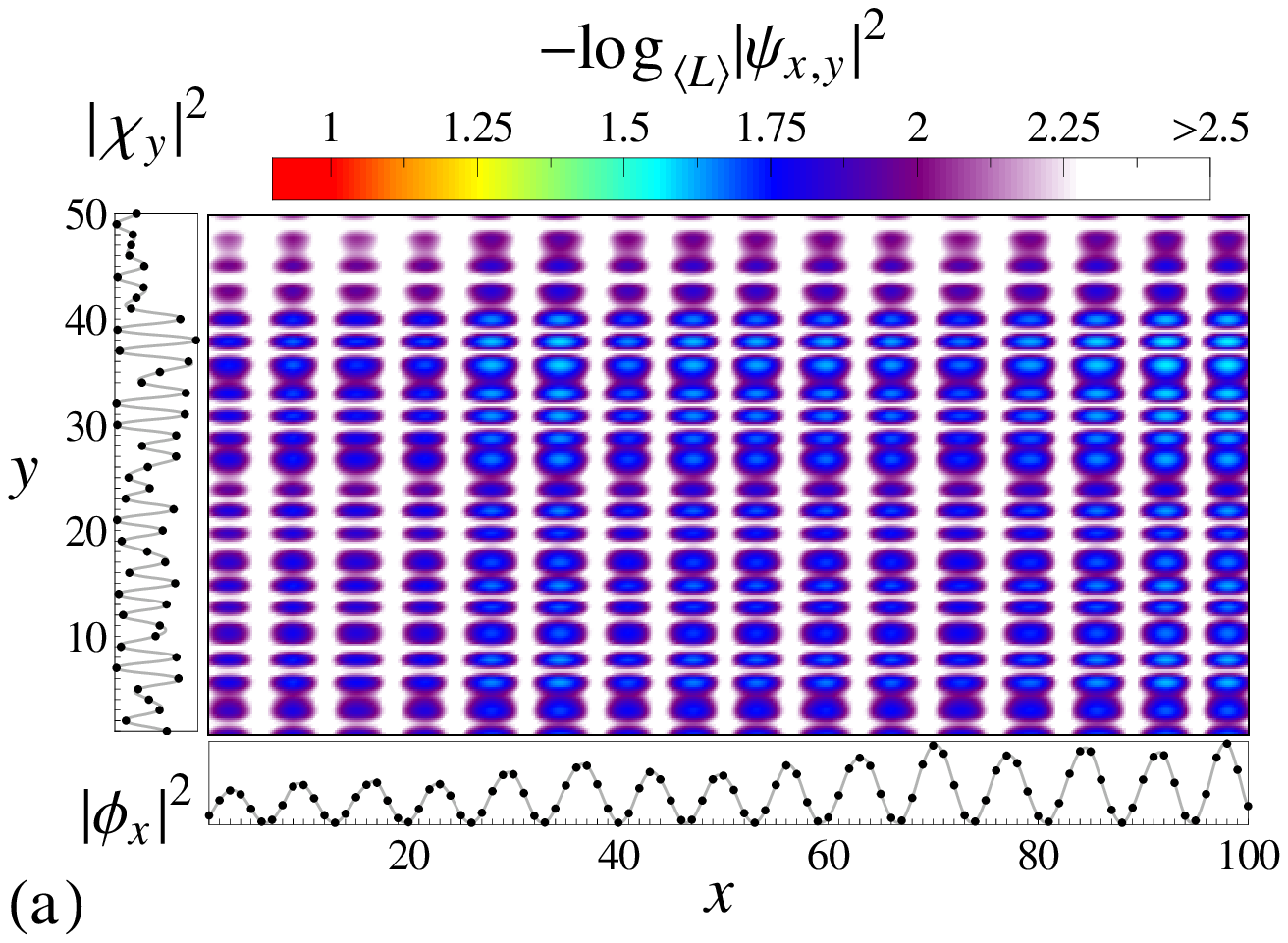}\hfill
  \includegraphics[width=\columnwidth]{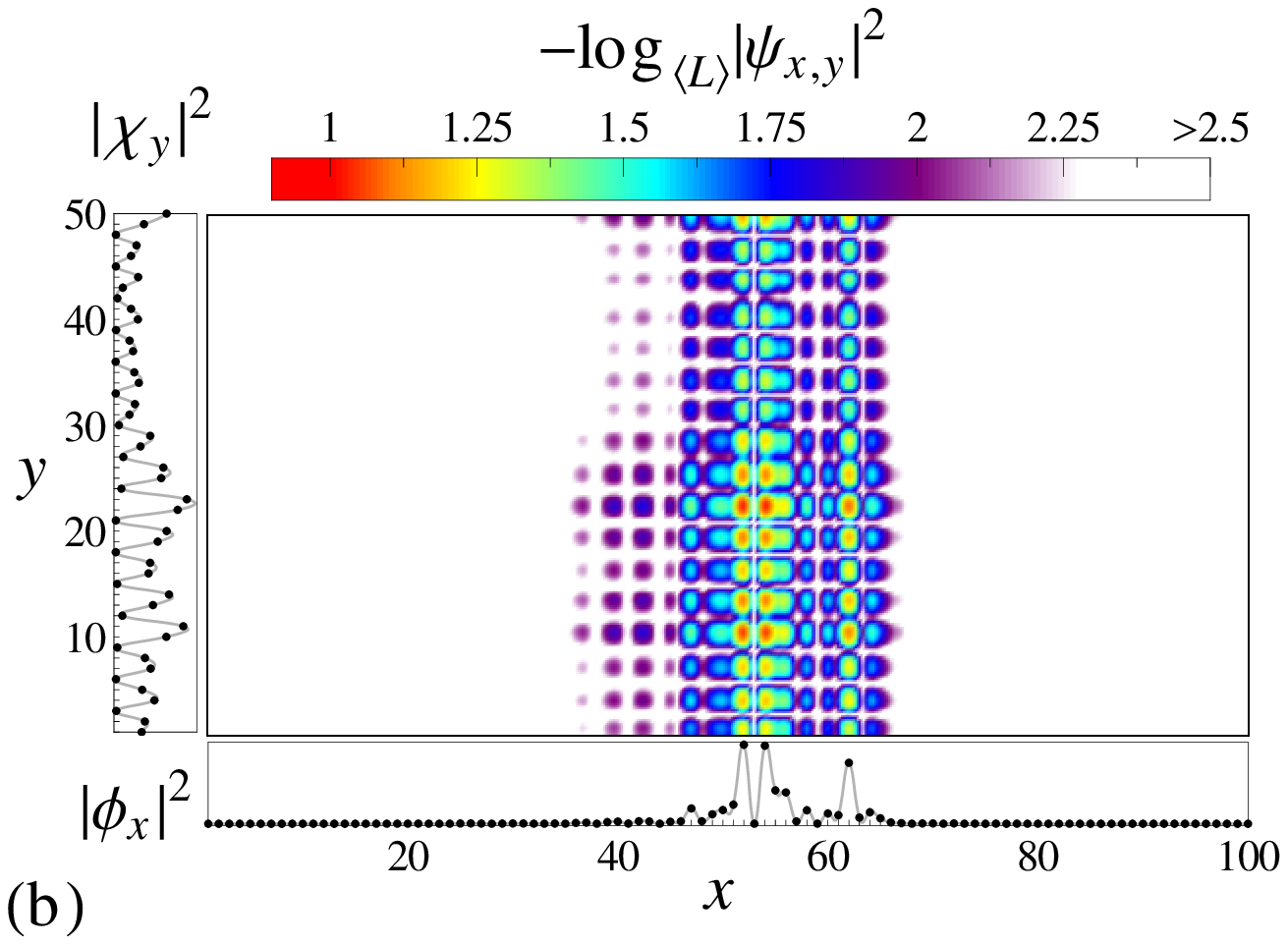}
 \caption{(Color online) Probability density $|\psi_{x,y}|^2$ of eigenstates of an $xy$-disordered system effectively delocalized in $y$ with (a) $\lambda_x>L_x$ and (b) $\lambda_x<L_x$. The parameters characterizing the potential are $m_\alpha=0$, $W_\alpha=3t$, $m_\beta=1$, $W_\beta=2$, $\xi_y\equiv\xi=2$ and eigenenergies  (a) $E/t= 1.7535$, (b) $E/t= -0.008711$. The states belong respectively to decoupled channels with (a) $p=0.044$ and (b) $p=-1.2029$. Notice that $\xi$ is chosen to satisfy Eqs.~\eqref{eq:resoTL} and \eqref{eq:delocaY} only.
 In both cases the system length and width are $L_x=100$ and $L_y=50$, respectively. The left and bottom panels show, respectively, the distributions $|\chi_y|^2$ and $|\phi_x|^2$ that construct the eigenstate according to \eqref{eq:2Dstate}. The color is determined by the value of $-\log_{\langle L\rangle}|\psi_{x,y}|^2$ where $\langle L\rangle\equiv\sqrt{L_xL_y}$ (in this scale, 2 is the average value of an extended state). A second order interpolation of the distributions is used to smooth the visualization.}
 \label{fig-delocaY}
\end{figure*}

Expression \eqref{eq:lly} can be used to estimate the value of $\xi$ that we need if we want the extended states that we have generated in the $x$-direction to be also \emph{effectively} delocalized ($\lambda_y>L_y$) in the vertical direction.
By using Eq.~\eqref{eq:lly} for $p=0$ we estimate that these states will be effectively delocalized in the $y$ direction when
\begin{equation}
 |\xi|> \frac{1}{2}\left[\frac{W_\beta^2 L_y}{24}+m_\beta^2\right]^{1/2}.
 \label{eq:delocaY}
\end{equation}
For the case $m_\beta=0$ we see that for systems a few hundred sites wide a value of $\xi\sim W_\beta$ will roughly ensure effective delocalization over the whole system of the resonant channel states, as shown in Fig.~\ref{fig-delocaY}(a).
Alternatively, we can also find effective delocalization in $y$, while keeping the localized character in the $x$ direction, as displayed in Fig.~\ref{fig-delocaY}(b).

Therefore, the localization properties of $xy$-disorder can be engineered almost at will, and independently in the $x$ and $y$ directions, giving rise to the 4 basic configurations shown in Figs.~\ref{fig-states} and \ref{fig-delocaY}.

\section{Engineering of transport and spectral properties}
\label{sec-engineering}
\subsection{Transport regimes in $xy$-disorder}
\label{sec-transport}
\begin{figure}
 \includegraphics[width=.95\columnwidth]{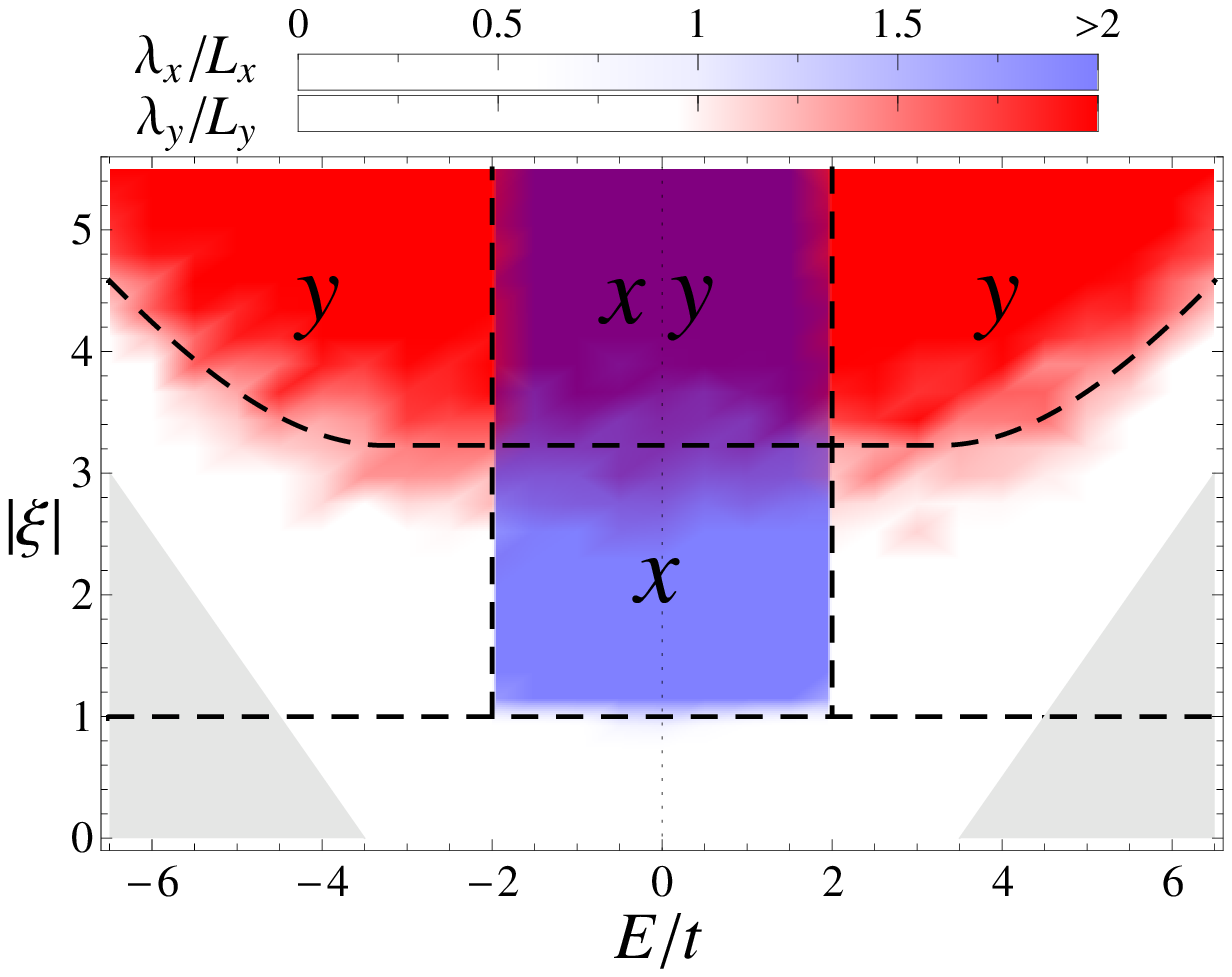}
 \caption{(Color online) Diagram of transport regimes for $|\xi|$ versus $E$ in a 2D system with $xy$-disorder of size $L_x=L_y=1000$ and parameters
 $W_\alpha=t$, $m_\alpha=0$, $W_\beta=1$, $m_\beta=2.5$. The dashed lines mark the limits of the efficient transport regimes obtained analytically and given by Eq.~\eqref{eq:resoTL} (straight horizontal dashed line), Eq.~\eqref{eq-tx2} (vertical dashed lines) and Eqs.~\eqref{eq-ty1} and \eqref{eq-ty2} (bent dashed line).
 The density plots in the background (color scales are indicated above the main panel) show numerically calculated localization lengths $\lambda_x$ (white-to-blue semitransparent) and $\lambda_y$ (white-to-red). The $(E,|\xi|)$ space was discretized in 675 points ($27\times25$) for the calculation, and the results averaged over 100 disorder realizations at every point. The density plot uses a linear interpolation to smooth the visualization. Labels $x$ and $y$ highlight regions where $\lambda_x>L_x$ and $\lambda_y>L_y$ respectively. The gray wedges at the lower corners correspond to regions outside the spectrum of the system.}
 \label{fig-tpd1}
\end{figure}

In Fig.~\ref{fig-tpd1} we show the diagram of transport regimes $|\xi|$ versus $E$ for a 2D system of size $L_x=L_y=1000$.
By comparing the magnitude of the localization length against the system size, we can distinguish regions of efficient transport in direction $x$, in $y$ or in both.
The lines separating the different phases are calculated 
in Appendix \ref{ap-transport}. We confirm the validity of the diagram by numerical calculations of localization lengths ---estimated from transmission probabilities in $x$ and $y$ with open boundary conditions--- averaged over 100 realizations of the system, displayed by the color density plots in Fig.~\ref{fig-tpd1}.

Transmission in the $x$ direction emerges from the existence of resonant or quasi-resonant channels, whereas in $y$ it is due to effectively delocalized states only. Thus it must be clear that in the thermodynamic limit, $L_x,L_y\rightarrow\infty$, only the efficient transport along $x$ remains. The diagram is system-size dependent with respect to the transport properties in the $y$-direction. Nevertheless, for finite systems, there exists the interesting possibility of realizing switching devices whose transmitting behaviour in $x$ and $y$ is highly controllable.

\begin{figure}
 \includegraphics[width=.97\columnwidth]{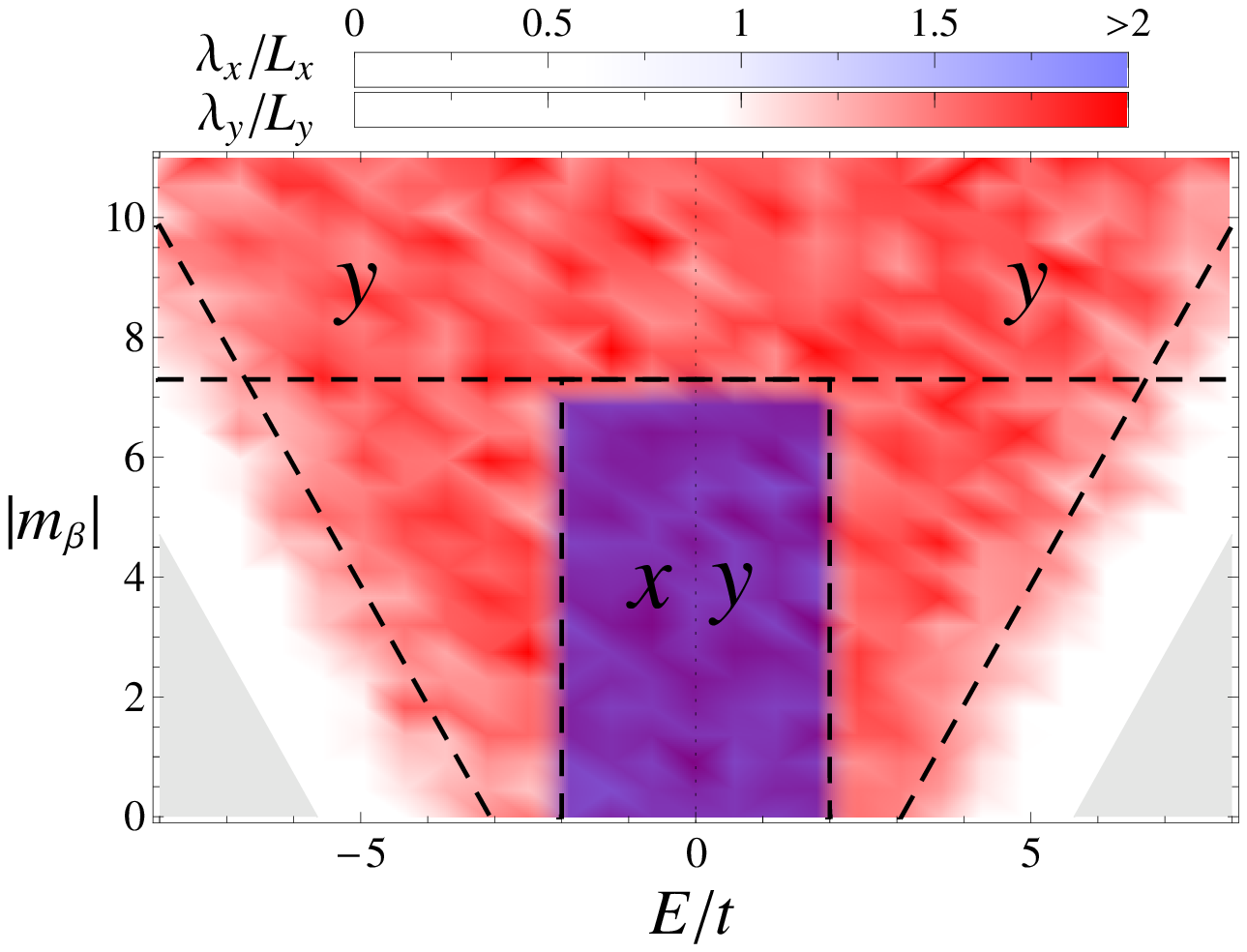}
 \caption{(Color online) Diagram of transport regimes for $m_\beta$ versus $E$ in a 2D system with $xy$-disorder of size $L_x=L_y=1000$ and parameters
 $W_\alpha=t$, $m_\alpha=0$, $W_\beta=1$, $\xi=3.4$. The dashed lines mark the limits of the efficient transport regimes and can be obtained from the equations in Appendix \ref{ap-transport}. The density plots in the background show numerical calculations of the localization lengths as in Fig.~\ref{fig-tpd1}.}
 \label{fig-tpd2}
\end{figure}

Different configurations of the diagram can be obtained by changing the disorder parameters $W_\alpha$, $m_\alpha$, $W_\beta$ and $m_\beta$. For example, $m_\beta$ controls the appearance of the resonant channels in $x$, as well as the width of the plateau region in the border of effective delocalization in $y$. On the other hand, $m_\alpha\neq0$ breaks the symmetry of the diagram, the region of efficient transport in $y$ shifts laterally preserving the plateau region and decaying asymmetrically on the sides. In Fig.~\ref{fig-tpd2}, we show an alternative diagram of $m_\beta$ vs $E$, where different transport regimes can be triggered simply by changing the mean value of the $\beta_y$ distribution, even for a fixed disorder realization.

\subsection{Coexistence of extended and localized states}
\label{sec-coexistence}
In a finite $xy$-system, the fulfilment of the resonance condition (as discussed in Section \ref{sec-zero}) induces $L_x$ Bloch states in the $x$ direction for
$E\in[-2t,2t]$. Nevertheless, the other $(L_y-1) L_x$ states coming from the remaining
decoupled channels will be localized in $x$, and can in principle have energies inside the range $[-2t,2t]$.
Therefore, in general, we should expect coexistence of extended and localized states in $[-2t,2t]$. The overlapping of the spectrum of resonant and localized channels can be seen in Fig.~\ref{fig-dosresonant}(a).
We have checked that the level spacing distribution in this region includes
that for Bloch states on a background of Poissonians corresponding to localized states, as also found in Ref.~\onlinecite{MouCL10} for a ladder model supporting coexistence.

As $L_x$ and $L_y$ grow, the range $[-2t,2t]$ will potentially become densely populated by extended and localized states. The filling of that energy interval by Bloch states in a continuous manner will render the presence of localized states irrelevant, as far as transport in the $x$ direction is concerned.

\subsection{Avoiding the coexistence of extended and localized states}
\label{sec-avoid}

The $c$-th decoupled channel is characterized by an on-site energy distribution with mean $p_cm_\alpha$ and width $|p_c|W_\alpha$, and whose spectral boundaries are given by Eq.~\eqref{eq-channelspectrum}. Thus $m_\alpha$ controls the position of the spectrum of the different channels.
The avoidance of coexistence is only possible if the $\alpha_x$'s have all the same sign, i.e.~$|m_\alpha| > W_\alpha/2$. Otherwise the spectrum of all decoupled channels always includes $E=0$, as the lower and upper bounds have opposite signs, thus giving rise to overlapping with the range $[-2t,2t]$.

For simplicity we consider $m_\alpha>0$. Then we can work with $|p_c|$ and consider positive energies only without loss of generality.
In order to avoid coexistence, we have to ensure that the lowest energy of all regions of localized states is greater than $2t$.
In view of Eq.~\eqref{eq-channelspectrum} this condition is written as
\begin{equation}
  \left(m_\alpha-\frac{W_\alpha}{2}\right)\min|p_c| \geqslant 4t,
 \label{eq:lap}
\end{equation}
where $\min|p_c|$ stands for the minimum non-vanishing $|p_c|$. Let us recall that the resonance condition is satisfied and thus zero is an eigenvalue of $\mathbf{P}$ [Eq.~\eqref{eq:p}].

Once the disorder realization of the system is generated, i.e.~fixed sequences $\{\alpha_x\}$, $\{\beta_y\}$ and $\xi$, we can obtain the value of $\min|p_c|$ and then change $m_\alpha$ to satisfy the inequality.
We note that \eqref{eq:lap} is formally independent of the system size, but only reasonable for moderate $L_y$. Since we are in the region of resonance, for large $L_y$ the $\min|p_c|$ will get arbitrarily small (Section \ref{sec-largesize}) and the required $m_\alpha$ to avoid overlapping will be very large.

\begin{figure}
  \includegraphics[width=.95\columnwidth]{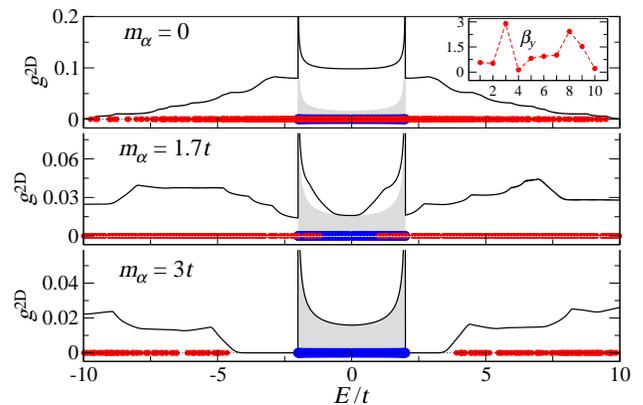}
 \caption{(Color online) Spectrum of $xy$-system with resonant channel and parameters $L_y=10$, $W_\alpha=2t$, $\xi=4.23 152$ for different $m_\alpha$. Black lines indicate DOS for $L_x=\infty$ as obtained from FEF, while blue (red) points on the $x$ axis mark eigenenergies of extended (localized) states for a finite realization with $L_x=100$. The shaded gray region highlights the contribution of the resonant channel to the total DOS. The inset shows the $\beta_y$ sequence considered characterized by $W_\beta\simeq 3$. According to Eq.~\eqref{eq:lap1} blue-red coexistence should be avoided for $m_\alpha\gtrsim3$.}
 \label{fig-over}
\end{figure}

It is also possible to obtain some general bounds for $m_\alpha$ without having to apply the above inequality to each particular realisation of the disorder.
For a fixed value of the width of the system $L_y$, we can estimate $\min|p_c|$ from the distribution $\varrho(p)$ around $p\sim 0$. A rough estimation is obtained by assuming a constant distribution, $\varrho=(W_\beta+4|\xi|)^{-1}$, from which it follows that $\min|p_c|\geqslant (W_\beta+4|\xi|)/L_y$.
Condition \eqref{eq:lap} translates into
\begin{equation}
 m_\alpha > \frac{4 t L_y}{W_\beta+4|\xi|} + \frac{W_\alpha}{2},
 \label{eq:lap1}
\end{equation}
which gives an approximation, usually overestimated,
of the minimum $m_\alpha$ value to avoid coexistence.
In Fig.~\ref{fig-over} we show examples of spectrum engineering of the overlapping between extended and localized states.

\section{Conclusions}
\label{sec-conclusions}

Our results show how a controlled choice of quite a simple, structured
disorder as given by Eqs.~\eqref{eq-disorder} and \eqref{eq-gamma} can lead
to extended states. By varying the parameters of the individual
disorder distributions, we can engineer the localization and transport properties of the system 
independently in each direction. We can furthermore control the energy 
coexistence of extended and localized states in the spectrum. 
Although our approach is based on the Anderson model, we expect our results to hold
in the many realms of Anderson localization physics in
general.\cite{HasSMI08,RicRMZ10,BilJZB08,CleVRS06,RoaDFF08,LemCSG09,WieBLR97,MooPYB08,FaeSPL09}

For an experimental realisation of our model, the robustness of our
results to small deviations away from the exact resonance conditions for quasi-1D systems,
as shown in Section \ref{sec-stability}, and 
the broadening of the resonance condition for 2D systems (Section \ref{sec-largesize}) 
is of course reassuring. 
Nevertheless, the disorder does require single-site control in order to adjust the
$\epsilon_{x,y}$ and $\gamma_{x,y}$ values as needed. 
The key feature expected of any experimental endeavour to measure our model is
henceforth the attempt to achieve full control by implementing single
site resolution for quantum state manipulation as well as quantum
state analysis. Fortunately, important progress in this direction has
already been achieved, e.g.\ in optical lattices.\cite{OspOWS06,GreF08,BakGPF09,WurLGK09,SimBMT11,ShrWEC10,WeiESC11,BloDN12}
A scheme combining a disordered optical potential and an additional
harmonic trap has been recently proposed \cite{PezS11} to observe
experimentally the coexistence of extended and localized
wavefunctions. We emphasize that our $xy$-disorder can provide a
genuine and controllable coexistence of extended and localized wavefunctions that does
not occur in usual disordered systems exhibiting delocalization
transitions.

While we have analysed 2D models for simplicity, the product structure for the disorder presented here  
can be straightforwardly extended to 3D systems, where similar features are to be expected, 
in particular, the possibility to engineer the transport properties in the three spatial directions independently. 

\begin{acknowledgments}%
We thank Kai Bongs for stimulating discussions and encouragement, and Andreas Buchleitner for a careful reading of the manuscript. 
We gratefully acknowledge EPSRC (EP/F32323/1, EP/C007042/1, EP/D065135/1)
for financial support. A.R.\ acknowledges financial support from the German DFG (BU 1337/5-1, BU 1337/8-1) and the Spanish MICINN (FIS2009-07880), and the hospitality of Departamento de F\'isica Fundamental at the University of Salamanca.
A.C.\ and R.A.R.\ are thankful to the Royal  Society for financial assistance (India-UK Science Networks). A.C.\ gratefully acknowledges the hospitality of the Centre for Scientific Computing in the Department of Physics, University of Warwick.
\end{acknowledgments}

\appendix

\section{Functional equation formalism}
\label{ap-FEF}
The DOS and the localization length of an infinite 1D disordered chain, can be calculated numerically using
the FEF.\cite{Rod06} This technique has been applied to obtain the
spectral properties in the thermodynamic limit of different quantum-wire models with correlated and uncorrelated disorder.\cite{RodCer06,CerRod05,CerRod03,CerRod02} Here we give a brief overview of the formalism applied to 1D tight-binding models with diagonal disorder, described by the equation $(E-\varepsilon_x)\phi_x=\phi_{x+1}-\phi_{x-1}$, where the on-site energies are obtained randomly from a continuous distribution $\mathcal{P}(\varepsilon)$. For such a system, the functional equation reads,
\begin{equation}
 \mathcal{W}(\theta) = \int d\varepsilon \mathcal{P}(\varepsilon) \mathcal{W}(T(\theta;E,\varepsilon)) -\mathcal{W}(\pi/2) +1,
 \label{eq:FE}
\end{equation}
where
$T(\theta;E,\varepsilon)=\arctan\left(E-\varepsilon - 1/\tan\theta \right)$,
with the conditions $\theta\in[0,\pi)$ and $\mathcal{W}(\theta+n\pi)=\mathcal{W}(\theta)+n$ for integer $n$.
The angular variable, $\theta$, follows from the representation of the wavefunction amplitude $\phi$ in polar coordinates. The function $\mathcal{W}(\theta)$ corresponds to the cumulative distribution function of $\theta$ in the disordered chain in the thermodynamic limit.
The DOS $g(E)$ and the Lyapunov exponent $\eta_x(E)$ as functions of the energy are written in terms of $\mathcal{W}(\theta)$ as,
\begin{align}
  g(E) =& \left|\frac{d\mathcal{W}(\pi/2)}{dE}\right|, \\
  \eta_x(E) =& \frac{1}{2}\int  d\varepsilon \mathcal{P}(\varepsilon) \ln f(\pi;E,\varepsilon) \notag\\
  &- \frac{1}{2}\int_0^{\pi} d\theta \mathcal{W}(\theta) \int d\varepsilon \mathcal{P}(\varepsilon) \frac{f^\prime(\theta;E,\varepsilon)}
  {f(\theta;E,\varepsilon)},
\end{align}
where $f(\theta,\varepsilon)=1-(E-\epsilon)\sin2\theta + (E-\varepsilon)^2\cos^2\theta$, and the prime indicates differentiation with respect to $\theta$. By solving numerically Eq.~\eqref{eq:FE} for different values of the energy, the spectral and localization properties of the infinite system can be obtained using the latter expressions.

\section{Transfer-matrix method}
\label{ap-tmm}

The transfer-matrix method (TMM) allows for a very memory efficient way to iteratively calculate the decay length 
of electronic states in a quasi-1D system of width $L_y$ for lengths $L_x \gg L_y$.\cite{KraM93} Equation (\ref{eq:canonical}) has to be rearranged into a form where the $\boldsymbol{\Psi}_{x+1}$ amplitudes of sites in layer $x+1$ --- when $x$ is chosen as the direction of transfer --- is calculated solely from parameters of sites in previous layers $x$ and $x-1$,
\begin{eqnarray}
 \boldsymbol{\Psi}_{x+1} &=& t^{-1}(E \openone -\boldsymbol{\epsilon}_x)\boldsymbol{\Psi}_x - \boldsymbol{\Psi}_{x-1} \label{eq-TMM_singles}\quad .
\end{eqnarray}
Equation \eqref{eq-TMM_singles} can be expressed in standard transfer-matrix form as
\begin{equation}
 \left[ \begin{array}{c}
   \boldsymbol{\Psi}_{x+1} \\ \boldsymbol{\Psi}_x
 \end{array} \right] =
\underbrace{
 \left[ \begin{array}{cc}
   t^{-1}(E\openone - \boldsymbol{\epsilon}_x) & -\openone \\ \openone &
\mathbf{0}
 \end{array} \right]}_{\mathbb{T}_n}
 \left[ \begin{array}{c}
   \boldsymbol{\Psi}_x \\ \boldsymbol{\Psi}_{x-1}
 \end{array} \right] \label{eq-TMM},
\end{equation}
where $\mathbf{0}$ and $\openone$ are the $L_y \times L_y$ zero and unit matrices, respectively.
Formally, the transfer matrix $\mathbb{T}_x$ is used to `transfer' electronic amplitudes $\boldsymbol{\Psi}$ from one slice to the next and repeated multiplication of this gives the global transfer matrix $\tau_{L_x} = \prod^{L_x}_{x=1} \mathbb{T}_x$. The limiting matrix $\Gamma \equiv \lim_{L_x\to\infty} \left(\tau_{L_x} \tau_{L_x}^\dagger \right)^{\frac{1}{2L_x}}$ exists\cite{Ose68,Arn98} and has eigenvalues $e^{\pm \eta_i}$, $i=1, \ldots, L_y$. The inverse of these Lyapunov exponents $\eta_i$ are estimates of decay/localization lengths and the physically most relevant largest decay length in the $x$ direction is
$\lambda_x =1/\min\eta_i$.

\section{Transport regimes}
\label{ap-transport}
The transport properties of the system along $x$ and $y$ can be engineered by tuning the parameters of the system: $m_\alpha$, $m_\beta$, $W_\alpha$, $W_\beta$, $\xi$ and $E$. A qualitative diagram of the transport regimes, for sufficiently large but finite $L_x$ and $L_y$, can be obtained by using the weak disorder expressions for the localization lengths $\lambda_x$ and $\lambda_y$. For simplicity we assume here that $m_\alpha, m_\beta \geqslant 0$ and we only focus on relations $|\xi|$ versus $E$.
\setcounter{paragraph}{0}
\paragraph{Transport along the $x$-direction:}
The basic condition to have resonant channels for large $L_y$ is given by Eq.~\eqref{eq:resoTL}.
Let us approximate that $\varrho(p)\equiv\varrho$ is constant for all $p$ and thus $\varrho=(W_\beta+4|\xi|)^{-1}$. Then for finite $L_y$, when the above condition is satisfied, we will find a value of $p$ as small as $|p_s|\leqslant(W_\beta+4|\xi|)/(2L_y)$. The maximum localization length in $x$ (provided by a channel with a disorder distribution of width $|p_s| W_\alpha\ll 1$ and mean $p_s m_\alpha\simeq0$), then satisfies
\begin{equation}
 \lambda_x \geqslant \frac{96 L_y^2 (4t^2-E^2)}{W_\alpha^2(W_\beta+4|\xi|)^2}, \qquad  E\in[-2t,2t].
\end{equation}
We can ensure efficient transport in $x$ by enforcing that the minimum bound for $\lambda_x$ is larger than $L_x$. This implies
\begin{equation}
 |\xi| \leqslant \frac{L_y}{W_\alpha} \sqrt{\frac{6(4t^2-E^2)}{L_x}}-\frac{W_\beta}{4}.
 \label{eq-tx2}
\end{equation}
This condition gives a continuum region of $\xi$ values providing transmitting behaviour in $x$. Note, however, that
$\lambda_x>L_x$ will also be achieved outside this region, around resonant values of $\xi$ ($p_s=0$).
\paragraph{Transport along the $y$-direction:}
The localization length in $y$ is determined by the localization properties of the eigenstates of the matrix $\mathbf{P}$. The localization length in $p$ is given by
\begin{equation}
 \lambda_y(p)= \frac{24}{W_\beta^2} \left[4\xi^2-\left(p -m_\beta\right)^2\right],
 \label{eq-llyap}
\end{equation}
for $p\in [- 2|\xi|+m_\beta, m_\beta + 2|\xi|]$ and $|\xi|\gg W_\beta/\sqrt{12}$. In this weak disorder approximation the localization length has a parabolic dependence in $p$ with its maximum value at $p=m_\beta$.
Since the localization length in $y$ is determined by the value of $p$, all states of a certain decoupled channel will exhibit the same $y$-spreading [see Eq.~\eqref{eq:decoupled}]. The spectrum of a channel characterized by $p_c$ is
\begin{equation}
E\in \left[-2t + p_c\left(m_\alpha-\frac{\sigma W_\alpha}{2}\right),2t+ p_c\left(m_\alpha+\frac{\sigma W_\alpha}{2}\right)\right],
\end{equation}
where $\sigma\equiv \text{sign}(p_c)$. For a given energy $E$, the maximum localization length in $y$, and thus the relevant for transport processes, is given by $\lambda_y(p_c)$ such that $p_c$ is the closest value to $m_\beta$ whose spectrum includes $E$.

For $p_c=m_\beta$ we get the maximum attainable value of $\lambda_y$ in the system, and thus
$\lambda_y(E)=96\xi^2/W_\beta^2$ for
\begin{equation}
  E\in\left[-2t + m_\beta\left(m_\alpha-\frac{W_\alpha}{2}\right),2t+ m_\beta\left(m_\alpha+\frac{W_\alpha}{2}\right)\right],
  \label{eq-plateau}
\end{equation}
since any other $p_c$ whose spectrum overlaps with this range will provide smaller values of $\lambda_y$. The localization length in $y$ thus displays a characteristic plateau structure [see Fig.~\ref{fig-dosgeneric}(b)] whose width is proportional to the parameter $m_\beta$.

For energies larger than the interval above, there is overlapping of the spectra of channels with $p>m_\beta$ (recall that we assume $m_\alpha, m_\beta\geqslant 0$). For a given energy $E$, the maximum $\lambda_y$ is attained for $p_c$ such that $E$ coincides with the upper edge of the spectrum of the $p_c$-channel, since this will be the value of $p$ closest to $m_\beta$. Therefore we can identify $p_c=(E-2t)/(m_\alpha+W_\alpha/2)$ and substitute in Eq.~\eqref{eq-llyap} to obtain $\lambda_y$ as function of the energy.

For energies smaller than the interval \eqref{eq-plateau}, it can be seen that the maximum $\lambda_y$ is attained for $p_c$ such that $E$ coincides with the lower edge of the spectrum of the $p_c$-channel. Thus $\lambda_y(E)$ is obtained from Eq.~\eqref{eq-llyap} for $p_c=(E+2t)/(m_\alpha-\sigma W_\alpha/2)$. The expression and its energy-range of validity depends on whether the contributing $p_c$ is positive ($\sigma=1$) or negative ($\sigma=-1$), as well on the sign of ($m_\alpha-W_\alpha/2$).

Therefore, from the plateau structure in the interval \eqref{eq-plateau}, the localization length in $y$ decays with $E$. For $m_\alpha=0$ the dependence of $\lambda_y(E)$ is symmetric around $E=0$. A non-vanishing $m_\alpha$ induces a shift in the position of the plateau and can lead to an asymmetric decay of $\lambda_y(E)$ from its maximum value, left and right from the plateau. These features can be observed in the numerical results shown in Fig.~\ref{fig-dosgeneric}(b).

In order to have efficient transport along $y$ we require $\lambda_y\geqslant L_y$.
Following the reasoning given above it is possible to write general relations for $|\xi|$ to ensure the latter condition.
In order to avoid cumbersome expressions, we show them here only for the case $m_\alpha=0$:
\begin{equation}
 |\xi| \geqslant \frac{W_\beta\sqrt{L_y}}{4\sqrt{6}},
 \quad  E\in\left[-2t -m_\beta \frac{W_\alpha}{2}, 2t+ m_\beta \frac{W_\alpha}{2}\right],
 \label{eq-ty1}
 \end{equation}
 and
 \begin{subequations}
 \begin{gather}
|\xi| \geqslant \frac{1}{2}\sqrt{ \frac{L_y W_\beta^2}{24} + \left(\frac{|E|-2t}{W_\alpha/2}-m_\beta\right)^2},\\
 \intertext{for}
  |E| \in\left[2t+m_\beta \frac{W_\alpha}{2}, 2t +(m_\beta+2|\xi|)\frac{W_\alpha}{2}\right].
 \label{eq-ty2}
 \end{gather}
 \end{subequations}
The latter relations are symmetric around $E=0$. As discussed above, a non-vanishing $m_\alpha$ will shift the plateau structure \eqref{eq-ty1},
and it can break the symmetric behaviour around it. The formulas above provide a qualitative understanding of the influence of the different parameters on the transport regimes in the $y$ direction.


\begin{thebibliography}{99}
\bibitem{AshM76}
N.~W. Ashcroft and N.~D. Mermin, {\em Solid State Physics} (Saunders College,
  New York, 1976).

\bibitem{Yab87}
E. Yablonovitch, Phys. Rev. Lett. {\bf 58},  2059  (1987).

\bibitem{Joh87}
S. John, Phys. Rev. Lett. {\bf 58},  2486  (1987).

\bibitem{DeeJT98}
F.~R. Montero~de Espinosa, E. Jim\'enez, and M. Torres, Phys. Rev. Lett. {\bf
  80},  1208  (1998).

\bibitem{VasDFH98}
J.~O. Vasseur, P.~A. Deymier, G. Frantziskonis, G. Hong, B. Djafari-Rouhani,
  and L. Dobrzynski, J. Phys.: Condens. Matter {\bf 10},  6051  (1998).

\bibitem{BarAA09}
I.~O. Barinov, A.~P. Alodzhants, and S.~M. Arakelyan, Quantum Electronics {\bf
  39},  685  (2009).

\bibitem{GroP08}
M. Grochol and C. Piermarocchi, Phys. Rev. B {\bf 78},  035323  (2008).

\bibitem{TaoSY07}
A. Tao, P. Sinsermsuksakul, and P. Yang, Nature Nanotechnology {\bf 2},  435
  (2007).

\bibitem{ChrESG07}
A. Christ, Y. Ekinci, H.~H. Solak, N.~A. Gippius, S.~G. Tikhodeev, and O.~J.~F.
  Martin, Phys. Rev. B {\bf 76},  201405  (2007).

\bibitem{HepG11}
S.~P. Hepplestone and G.~P. Srivastava, Phys. Rev. B {\bf 84},  115326  (2011).

\bibitem{Min11}
S. Minardi, Mon. Not. R. Astron. Soc. 422, 2656 (2012).

\bibitem{RebWZI12}
J. Reboud, R. Wilson, Y. Zhang, M.~H. Ismail, Y. Bourquin, and J.~M. Cooper,
  Lab Chip {\bf 12},  1268  (2012).

\bibitem{And58}
P.~W. Anderson, Phys. Rev. {\bf 109},  1492  (1958).

\bibitem{KraM93}
B. Kramer and A. MacKinnon, Rep. Prog. Phys. {\bf 56},  1469  (1993).

\bibitem{EveM08}
F. Evers and A.~D. Mirlin, Rev. Mod. Phys. {\bf 80},  1355  (2008).

\bibitem{AbrALR79}
E. Abrahams, P.~W. Anderson, D.~C. Licciardello, and T.~V. Ramakrishnan, Phys.
  Rev. Lett. {\bf 42},  673  (1979).

\bibitem{DunWP90}
D.~H. Dunlap, H.-L. Wu, and P.~W. Phillips, Phys. Rev. Lett. {\bf 65},  88
  (1990).

\bibitem{BelBHT99}
V. Bellani, E. Diez, R. Hey, L. Toni, L. Tarricone, G.~B. Parravicini, F.
  Dom\'inguez-Adame, and R. G\'omez-Alcal\'a, Phys. Rev. Lett. {\bf 82},  2159
  (1999).

\bibitem{ZhaU04}
W. Zhang and S.~E. Ulloa, Phys. Rev. B {\bf 69},  153203  (2004).

\bibitem{SedKS11}
T.~A. Sedrakyan, J.~P. Kestner, and S. Das~Sarma, Phys. Rev. A {\bf 84},
  053621  (2011).

\bibitem{AubA80}
S. Aubry and G. {Andr\'e}, Ann. Israel Phys. Soc. {\bf 3},  133  (1980).

\bibitem{DemL98}
F.~A. B.~F. de~Moura and M.~L. Lyra, Phys. Rev. Lett. {\bf 81},  3735  (1998).

\bibitem{IzrK99}
F.~M. Izrailev and A.~A. Krokhin, Phys. Rev. Lett. {\bf 82},  4062  (1999).

\bibitem{KuhIKS00}
U. Kuhl, F.~M. Izrailev, A.~A. Krokhin, and H.-J. Stockmann, Appl. Phys. 
  Lett. {\bf 77},  633   (2000).

\bibitem{MouCLR04}
F.~A. B.~F. de~Moura, M.~D. Coutinho-Filho, M.~L. Lyra, and E.~P. Raposo,
  Europhys. Lett. {\bf 66},  585  (2004).

\bibitem{GuoX11}
A.-M. Guo and S.-J. Xiong, Phys. Rev. B {\bf 83},  245108  (2011).

\bibitem{SchWB11}
T. Scholak, T. Wellens, and A. Buchleitner, Europhys. Lett. {\bf 96},
   10001  (2011).

\bibitem{SchWB11b}
T. Scholak, T. Wellens, and A. Buchleitner, J. Phys. B: At. 
  Mol. Opt. Phys. {\bf 44},  184012  (2011).

\bibitem{Hil03}
M. Hilke,  Phys. Rev. Lett. {\bf 91}, 226403 (2003).

\bibitem{XioKE98}
S-J. Xiong, G. N. Katomeris, S. N. Evangelou, Ann. Phys. (Leipzig) {\bf 7}, 363 (1998)

\bibitem{MouD08}
F. {de Moura} and F. Dom\'inguez-Adame, Eur. Phys. J. B {\bf 66}, 165 (2008).

\bibitem{NdaRS04}
M.~L. Ndawana, R.~A. {R\"{o}mer}, and M. Schreiber, Europhys. Lett. {\bf 68},
  678  (2004).

\bibitem{AkaASN03}
Y. Akahane, T. Asano, B.-S. Song, and S. Noda, Nature {\bf 425},  944  (2003).

\bibitem{TonVOJ08}
C. Toninelli, E. Vekris, G.~A. Ozin, S. John, and D.~S. Wiersma, Phys. Rev.
  Lett. {\bf 101},  123901  (2008).

\bibitem{GarSTL11}
P.~D. Garc\'ia, R. Sapienza, C. Toninelli, C. L\'opez, and D.~S. Wiersma, Phys.
  Rev. A {\bf 84},  023813  (2011).

\bibitem{EstCBU12}
H. Estrada, P. Candelas, F. Belmar, A. Uris, F.~J. Garc\'ia~de Abajo, and F.
  Meseguer, Phys. Rev. B {\bf 85},  174301  (2012).

\bibitem{SanL10}
L. Sanchez-Palencia and M. Lewenstein, Nat. Phys. {\bf 6},  87  (2010).

\bibitem{BilJZB08}
J. Billy, V. Josse, Z. Zuo, A. Bernard, B. Hambrecht, P. Lugan, D. Clement, L.
  Sanchez-Palencia, P. Bouyer, and A. Aspect, Nature {\bf 453},  891  (2008).

\bibitem{RoaDFF08}
G. Roati, C. D'Errico, L. Fallani, M. Fattori, C. Fort, M. Zaccanti, G.
  Modugno, M. Modugno, and M. Inguscio, Nature {\bf 453},  895  (2008).

\bibitem{KonMZD11}
S.~S. Kondov, W.~R. McGehee, J.~J. Zirbel, and B. DeMarco, Science {\bf 334},
  66  (2011).

\bibitem{JenBMC12}
F. Jendrzejewski, A. Bernard, K. Muller, P. Cheinet, V. Josse, M. Piraud, L.
  Pezze, L. Sanchez-Palencia, A. Aspect, and P. Bouyer, Nat. Phys. {\bf 8},
  398  (2012).

\bibitem{KloRT05}
D.~K. Klotsa, R.~A. {R\"{o}mer}, and M.~S. Turner, Biophys. J. {\bf 89},  2187
  (2005).

\bibitem{ZhaYZD10}
W. Zhang, R. Yang, Y. Zhao, S. Duan, P. Zhang, and S.~E. Ulloa, Phys. Rev. B
  {\bf 81},  214202  (2010).

\bibitem{MouCL10}
F.~A. B.~F. de~Moura, R.~A. Caetano, and M.~L. Lyra, Phys. Rev. B {\bf 81},
  125104  (2010).

\bibitem{SilMC08}
S. Sil, S.~K. Maiti, and A. Chakrabarti, Phys. Rev. Lett. {\bf 101},  076803
  (2008).

\bibitem{SilMC08b}
S. Sil, S.~K. Maiti, and A. Chakrabarti, Phys. Rev. B {\bf 78},  113103
  (2008).

\bibitem{OspOWS06}
S. Ospelkaus, C. Ospelkaus, O. Wille, M. Succo, P. Ernst, K. Sengstock, and K.
  Bongs, Phys. Rev. Lett. {\bf 96},  180403  (2006).

\bibitem{GreF08}
M. Greiner and S. F{\"o}lling, Nature {\bf 453},  736  (2008).

\bibitem{BakGPF09}
W.~S. Bakr, J.~I. Gillen, A. Peng, S. F{\"o}lling, and M. Greiner, Nature {\bf
  462},  74  (2009).

\bibitem{WurLGK09}
P. W\"urtz, T. Langen, T. Gericke, A. Koglbauer, and H. Ott, Phys. Rev. Lett.
  {\bf 103},  080404  (2009).

\bibitem{SimBMT11}
J. Simon, W.~S. Bakr, R. Ma, M.~E. Tai, P.~M. Preiss, and M. Greiner, Nature
  {\bf 472},  307  (2011).

\bibitem{ShrWEC10}
J.~F. Sherson, C. Weitenberg, M. Endres, M. Cheneau, I. Bloch, and S. Kuhr,
  Nature {\bf 467},  68  (2010).

\bibitem{WeiESC11}
C. Weitenberg, M. Endres, J.~F. Sherson, M. Cheneau, P. Schau{\ss}, T.
  Fukuhara, I. Bloch, and S. Kuhr, Nature {\bf 471},  319  (2011).

\bibitem{BloDN12}
I. Bloch, J. Dalibard, and S. Nascimbene, Nat. Phys. {\bf 8},  267  (2012).

\bibitem{Rod06}
A. Rodr\'iguez, J. Phys. A: Math. Gen. {\bf 39},  14303  (2006).

\bibitem{Eco90}
E.~N. Economou, {\em Green's Functions in Quantum Physics} (Springer-Verlag,
  Berlin, 1990).

\bibitem{RomS04}
R.~A. R\"omer and H. Schulz-Baldes, Europhys. Lett. {\bf 68},  247  (2004).

\bibitem{foot-realsolution} While not all solutions of $\det\mathbf{P}(\xi)=0$ correspond to real $\xi$ values, this does not represent a problem. If $L_y$ is very small, the values of $\beta_y$ may be tuned to ensure real solutions, but in our experience for $L_y>5$ some real solutions are found without further restrictions on the values of $\beta_y$. Nevertheless, the $\beta$ distribution can be shifted by changing $m_\beta$ to ensure $\beta_y>0$, in which case all resonant $\xi$ values are real.

\bibitem{foot-precision} Throughout the text, we restrict ourselves to giving no more than $5$ significant digits for these resonant $\xi$ values. 

\bibitem{foot-bloch} The periodic eigenstates given by Eq.~\eqref{eq-bloch} become strictly Bloch states as $L_x\rightarrow\infty$ or if periodic boundary conditions are assumed for finite $L_x$.

\bibitem{Ger31}
S. Gersgorin, Izv. Akad. Nauk SSSR, Otd. Mat. Estest. Nauk, VII. Ser. No.6, 749 (1931).

\bibitem{GolL96}
G.~H. Golub and C.~F.~v. Loan, {\em Matrix Computations}, 3rd ed. (Johns
  Hopkins University Press, Baltimore and London, 1996).

\bibitem{Tho72}
D.~J. Thouless, J. Phys. C {\bf 5},  77  (1972).

\bibitem{HasSMI08}
K. Hashimoto, C. Sohrmann, M. Morgenstern, T. Inaoka, J. Wiebe, R.~A. R\"omer,
  Y. Hirayama, and R. Wiesendanger, Phys. Rev. Lett. {\bf 101},  256802
  (2008).

\bibitem{RicRMZ10}
A. Richardella, P. Roushan, S. Mack, B. Zhou, D.~A. Huse, D.~D. Awschalom, and
  A. Yazdani, Science {\bf 327},  665  (2010).

\bibitem{CleVRS06}
D. Cl{\'e}ment, A.~F. Var{\'o}n, J.~A. Retter, L. Sanchez-Palencia, A. Aspect,
  and P. Bouyer, New Journal of Physics {\bf 8},  165  (2006).

\bibitem{LemCSG09}
G. Lemari\'e, J. Chab\'e, P. Szriftgiser, J.~C. Garreau, B. Gr\'emaud, and D.
  Delande, Phys. Rev. A {\bf 80},  043626  (2009).

\bibitem{WieBLR97}
D.~S. Wiersma, P. Bartolini, A. Lagendjik, and R. Righini, Nature {\bf 390},
  671  (1997).

\bibitem{MooPYB08}
S.-H.~Y. Shayan~Mookherjea, Jung S.~Park and P.~R. Bandaru, Nature Photonics
  {\bf 2},  90  (2008).

\bibitem{FaeSPL09}
S. Faez, A. Strybulevych, J.~H. Page, A. Lagendijk, and B.~A. van Tiggelen,
  Phys. Rev. Lett. {\bf 103},  155703  (2009).

\bibitem{PezS11}
L. Pezz\'e and L. Sanchez-Palencia, Phys. Rev. Lett. {\bf 106},  040601
  (2011).

\bibitem{RodCer06}
A. Rodr\'iguez and J.~M. Cerver\'o, Phys. Rev. B {\bf 74},  104201  (2006).

\bibitem{CerRod05}
A. Rodr\'iguez and J.~M. Cerver\'o, Phys. Rev. B {\bf 72},  193312  (2005).

\bibitem{CerRod03}
J.~M. Cerver\'o and A. Rodr\'iguez, Eur. Phys. J. B {\bf 32},  537  (2003).

\bibitem{CerRod02}
J.~M. Cerver\'o and A. Rodr\'iguez, Eur. Phys. J. B {\bf 30},  239  (2002).

\bibitem{Ose68}
V.~I. Oseledets, Trans. Moscow Math. Soc. {\bf 19},  197  (1968).

\bibitem{Arn98}
L. Arnold, {\em Random Dynamical Systems} (Springer-Verlag, Berlin, 1998).

\end{thebibliography}


\end{document}